\documentclass[12pt]{report}          
\pdfoutput=1
\usepackage[intlimits]{amsmath}
\usepackage{amsfonts,amssymb}
\DeclareSymbolFontAlphabet{\mathbb}{AMSb}
\usepackage{apalike}
\usepackage{float,subfigure}
\usepackage[bf]{caption2}       
\setcaptionmargin{0.5in}
\usepackage{fancyheadings,fancybox,ifthen}
\usepackage{bu_ece_thesis}
\usepackage{url}
\usepackage{lscape,afterpage}
\usepackage{xspace}
\usepackage[table]{xcolor}

\usepackage[title,titletoc]{appendix}

\usepackage[dvips]{graphicx}

\begin{document}



\title{New Approaches to identify Gene-by-Gene Interactions in Genome Wide Association Studies}


\author{Chen Lu}

\degree=2

\prevdegrees{M.S., Northeastern University, 2007 \\ B.S., Beijing Institute of Technology, 2005}

\department{Department of Biostatistics}

\defenseyear{2013}
\degreeyear{2013}

\reader{First}{Jos{\'e}e Dupuis} {Professor of Biostatistics}
\reader{Second}{Eric D. Kolaczyk}{Professor of Mathematics \& Statistics }
\reader{Third}{Ching-Ti Liu}{Assistant Professor of Biostatistics}


\majorprof{Jos{\'e}e Dupuis}{\mbox{Professor of Biostatistics} \mbox{}}




\maketitle
\cleardoublepage


\approvalpage
\cleardoublepage




\newpage
\section*{\centerline{Acknowledgments}}
I would never have been able to finish my doctoral thesis without the guidance of my committee members, and the support and help from my family and friends.

First I would like to express my deepest appreciation to my major thesis advisor Professor Jos{\'e}e Dupuis, for her excellent guidance, thoughtfulness, immense knowledge and long-time support. I would like to sincerely thank my second advisor Professor Eric D. Kolaczyk, who guided me by his sharp thinking and extremely helpful suggestions from the very beginning of my research to writing up the papers. I would like to thank other committee members. Professor Ching-Ti Liu has provided me valuable suggestions in many aspects, including my final presentation. I would like to thank the committee chair Professor Kathryn L. Lunetta for her insightful suggestions and organizing the defense. I also would like to thank Professor Paola Sebastiani for her scientific advice and helpful discussions and suggestions.

I would like to thank my parents and my husband. They have been supportive and standing by me through the good times and bad. 

I also need to thank my friends Dr. Wei Vivian Zhuang, Dr. Ke Wang, Dr. Han Chen, Jacqui Milton and Shuai Wang for their constant supports and various discussions.
\cleardoublepage


\begin{abstractpage}

Genetic variants identified to date by genome-wide association studies only explain a small fraction of total heritability. Gene-by-gene interaction is one important potential source of unexplained heritability. In the first part of this dissertation, a novel approach to detect such interactions is proposed. This approach utilizes penalized regression and sparse estimation principles, and incorporates outside biological knowledge through a network-based penalty. The method is tested on simulated data under various scenarios. Simulations show that with reasonable outside biological knowledge, the new method performs noticeably better than current stage-wise strategies in finding true interactions, especially when the marginal strength of main effects is weak.

The proposed method is designed for single-cohort analyses.  However, it is generally acknowledged that only multi-cohort analyses have sufficient power to uncover genes and gene-by-gene interactions with moderate effects on traits, such as likely underlie complex diseases.   Multi-cohort, meta-analysis approaches for penalized regressions are developed and investigated in the second part of this dissertation. Specifically, I propose two different ways of utilizing data-splitting principles in multi-cohort settings and develop three procedures to conduct meta-analysis. Using the method developed in the first part of this dissertation as an example of penalized regressions, three proposed meta-analysis procedures are compared to mega-analysis using a simulation study. The results suggest that the best approach is to split the participating cohorts into two groups, to perform variable selection for each cohort in the first group, to fit regular regression model on the union of selected variables for each cohort in the second group, and lastly to conduct a meta-analysis across cohorts in the second group.

In the last part of this dissertation, the novel method developed in the first part is applied to the Framingham Heart Study measures on total plasma Immunoglobulin E (IgE) concentrations, C-reactive protein levels, and Fasting Glucose. The effect of incorporating various sources of biological information on the ability to detect gene-gene interaction is explored. For IgE, for example, a number of potentially interesting interactions are identified. Some of these interactions involve pairs in human leukocyte antigen genes, which encode proteins that are the key regulators of the immune response. The remaining interactions are among genes previously found to be associated with IgE as main effects. Identification of these interactions may provide new insights into the genetic basis and mechanisms of atopic diseases.

\end{abstractpage}
\cleardoublepage


\tableofcontents
\cleardoublepage

\listoftables
\cleardoublepage

\newpage
\listoffigures
\cleardoublepage

\chapter*{List of Abbreviations}
\begin{center}
  \begin{tabular}{lll}
    \hspace*{2em} & \hspace*{1in} & \hspace*{4.5in} \\
    CRP  & \dotfill & C-reactive Protein \\
    IgE  & \dotfill &  Immunoglobulin E\\
    GWAS  & \dotfill & Genome Wide Association Study \\
    KEGG   & \dotfill & Kyoto Encyclopedia of Genes and Genomes \\
    KKT & \dotfill & Karush-Kuhn-Tucker \\
    LD   & \dotfill & Linkage Disequilibrium \\
    MAF   & \dotfill & Minor Allele Frequency \\
    SE & \dotfill & Standard Error \\
    SNP   & \dotfill & Single Nucleotide Polymorphism \\
  \end{tabular}
\end{center}
\cleardoublepage


\newpage
\endofprelim
        
\cleardoublepage

\chapter{Introduction}
\label{chapter:Introduction}
\thispagestyle{myheadings}

Unlike Mendelian diseases, in which disease phenotypes are largely driven by mutation in a single gene locus, complex disease and traits are associated with a number of factors, both genetic and environmental, as well as lifestyle. In addition, while most Mendelian diseases are rare, many complex diseases are frightfully common, from asthma to heart disease, hypertension to Alzheimer's, and Parkinson's to various forms of cancer.

Arguably motivated by classical successes with Mendelian diseases and traits, the study of complex diseases and traits in the modern genomics era has focused largely on the identification of individually important genes. Genome-wide association studies (GWAS), the current state of the art, have been central to the discovery of many genes in various diseases (e.g., \cite{hindorff2010catalog}). However, unfortunately, the vast majority of genetic variants associated with complex traits identified to date explain only a very small amount of the overall variance of the trait in the underlying population \cite{manolio2009finding}. As a result, most GWAS findings thus far have had little clinical impact.

Currently, most GWAS are carried out one single nucleotide polymorphism (SNP) at a time. Typically, for each SNP a model is specified, relating disease status or disease trait to the SNP plus other potentially relevant covariates. The statistical significance of each SNP is quantified through the p-value of an appropriate test. Finally, a multiple testing correction is applied to correct the collection of p-values across SNPs. The end result is a list of SNPs declared to be significantly associated with the disease status or trait of interest, which in turn can be mapped to their closest genes, although some associations have been found in `gene deserts' \cite{hindorff2010catalog}. The single-SNP approach has the important attribute that it is (relatively) computationally efficient. But it can be severely under-powered because of the small effect size of most genetic variants identified to date \cite{hindorff2010catalog,manolio2009finding}. Additionally, this approach does not adjust for correlation among SNPs, nor does it extend in a natural manner to search for interactions between markers. In contrast, multiple regression (i.e., where multiple SNPs are modeled simultaneously) is a natural alternative. But naive implementation (i.e., incorporating all SNPs of interest) is both infeasible and undesirable. This is due to various reasons, including the sheer number of SNPs typically available (e.g.,  hundreds of thousands to millions), the comparatively small number of SNPs likely to be associated, and `small n, large p' problems.

Recently, however, computationally efficient multiple regression strategies for GWAS have begun to emerge that employ various methods of high-dimensional variable selection (e.g.,\cite{wu2009genome,ma2010identification,wu2010screen,szymczak2009machine,logsdon2010variational,ayers2010snp,zhou2010association}).
Compared to traditional single-SNP methods, penalized regression methods have been found to yield fewer correlated SNPs \cite{ayers2010snp} and to be capable of producing substantially more power while having a lower false discovery rate \cite{he2011variable}. Furthermore, regression methods can include SNP by SNP interactions in a natural manner.  However, to date this typically has been done in a greedy, stage-wise fashion, by fitting main effect models first and then restricting attention to interactions among those effects found significant \cite{wu2009genome,wu2010screen}. In addition, the above work makes limited or no use of supplementary biological information on, for example, biological pathways and gene function.

We propose a novel network-guided statistical methodology to facilitate the discovery of gene by gene (GxG) interactions associated with complex quantitative traits related to human disease, one which addresses both of the short-comings cited above.  Main effects and interaction effects in our model are chosen simultaneously, thus allowing for the possibility of detecting genes for which the marginal main effect is weak.  Variable selection is done through penalized regression using sparse estimation principles.  The penalty allows for the incorporation of information on biological pathways and gene function into the analysis of continuous traits related to human disease.  In doing so, this penalty acts as an informal prior distribution on the set of possible GxG interactions, which in practice allows the investigator to reduce the number of interactions examined for the model from the nominal and computationally prohibitive $O((\#\hbox{ of SNPs})^2)$ to a more manageable, say,  $O(\#\hbox{ of SNPs})$.

In Chapter 2, we describe our statistical approach. We introduce the model and our proposed penalty, describe how biological information is incorporated into the penalty and explain the optimization algorithm used for model fitting and a strategy for choosing tuning parameters. The design and results of an extensive simulation study are also presented in Chapter 2. We examine models with varying degrees of interactions and penalties reflecting different extents of biological knowledge. Simulations indicate that, given relevant pathway information, our approach performs well in finding true interactions without losing the ability of detecting main effects, and can noticeably outperform existing stage-wise methods. 

In Chapter 3, we extend our method to the multi-cohort setting. We test and compare four meta-analysis approaches in simulations, namely procedures A, B, C and D. Procedure A is the ideal situation that we pool individual data from different cohorts together and apply our proposed method to the combined dataset. This is often not possible in practice due to issues arising from patients\rq{} confidentiality. But it is a \lq{}gold standard\rq{} to which other approaches can be compared. Procedure B consists of splitting cohorts into two groups, performing variable selection on the cohorts in the first group, conducting regression on the union list of selected variables using the cohorts in the second group and meta-analyzing the results from the regression using the second group. Procedure C consists of splitting each cohort into two parts, performing variable selection using our new method on one part and regressing on selected terms on the other part, and conducting meta-analysis using result from regression across all cohorts. Procedure D is a variation of procedure C. In procedure C, cohorts may regress on different lists of selected variables. However, in procedure D, we combine all selected terms across cohorts after we perform variable selection using our new method on the first half of data and regress on the union list using the second half data for all cohorts. In Chapter 3, simulation studies suggests that procedure B is the one that performs most closely to procedure A and thus the one that should be used when conducting meta-analysis using our proposed methodology in Chapter 2.

In Chapter 4, we apply our proposed methodology to three real data examples in Framingham Heart Study. We investigated gene by gene interactions along with main effects for the following traits: total plasma Immunoglobulin E (IgE) concentrations, C-reactive protein (CRP) levels and Fasting Glucose. The analyses are performed on SNPs with low linkage disequilibrium (LD). The performance of our method under moderate LD is also assessed. Different outside biological information (i.e. pathway databases) are incorporated into the analyses. We also apply the stage-wise method for interaction investigation, to compare with the results of our method. In Chapter 4, we identify some interesting interactions that may be biological meaningful.

In Chapter 5, we conclude this dissertation with some additional discussion and potential future research related to this work.


\cleardoublepage

\chapter{Network-Guided Sparse Regression Modeling for Detection of Gene by Gene Interactions}
\label{chapter:chp2}
\thispagestyle{myheadings}



In this chapter, we introduce a novel statistical methodology to detect gene-by-gene interactions. This method utilizes penalized regression and sparse estimation principles, and incorporates outside biological knowledge though a network-based penalty. We will present the performance of this method under various scenarios in simulation studies and show that this method outperforms stage-wise strategies.

\section{Methods}
\label{sec:Methods}

\subsection{Modeling Gene by Gene Interaction}
\label{subsec:Modeling}

Let Y be a quantitative trait of interest, and let $\{X_j\}_{j=1}^p$ be $p$ predictors representing SNPs.
To include interactions, we are interested in a model of the form
\begin{equation}
Y = \beta_0 + \sum_{j=1}^p \beta_j X_j  + \sum_{k>j} \beta_{jk} X_{jk} + \epsilon \enskip \label{eq:01}
\end{equation}
where $X_{jk}=X_j X_k$. We expect that both the $\beta_j$'s and the $\beta_{jk}$'s are sparse, since it is unlikely that there is more than a small fraction of SNPs affecting the phenotype Y, either as main effects or as interactors.

In practice, $p$ will range from hundreds to millions. Our goal is to fit the high dimensional model (\ref{eq:01}) to data. When $p$ is large but only a small percentage of predictors and interactions are present in the true model, a general approach is to minimize a penalized regression criterion. Accordingly, we propose to estimate the coefficients $\mathbb{\beta} = (\{\beta_j\}, \{\beta_{jk}\})^T$ in our model using a penalized least-squares criterion.  Let $\mathbf{Y}=(Y_1,\ldots,Y_n)^T$, $\mathbf{X}_j=(X_{1j},\ldots,X_{nj})^T$ and
$\mathbf{X}_{jk} = (X_{1j}X_{1k}, \ldots, X_{nj}X_{nk})^T$  represent our variables $Y$, $X_j$, and $X_{jk}$ collected
over $n$ samples.  Our criterion is then written
\begin{equation}
\mathbb{\tilde{\beta}} = \arg\min_{\mathbb{\beta}}
	\frac{1}{2}\big|\big| \textbf{Y} - \sum_{j=1}^p \beta_j \textbf{X}_j - \sum_{k>j} \beta_{jk} \textbf{X}_{jk}\big|\big|^2
+ P_W(\mathbb{\beta})\enskip .    \label{eq:06}
\end{equation}

Penalized linear regression has been found to be a powerful tool for fitting high-dimensional models, particularly in situations where the nominal number of variables is large relative to the number of observations (e.g \cite{buhlmann2011statistics}).  In the context of GWAS, typically $p\gg n$.  Hence, it is impossible to fit a model with the full set of $O(p^2)$ nominal interactions among all $p$ SNPs.  However, the coefficient vector $\mathbb{\beta}$ is expected to be sparse.  Therefore, a penalty function that enforces sparseness can be helpful here, by encouraging the optimization in (\ref{eq:06}) to find solutions in which a large percentage of the main effects and their interactions are zero, thus dropping the corresponding terms from the model.

Following standard practice, we wish to include interactions only if their corresponding main effects are also included in the model.  The construction of the sparseness penalty $P_W$ therefore must be handled with some care, so as to enforce the resulting hierarchical constraint among coefficients .  In addition, we would like our penalty to allow for the use of biological knowledge (e.g., biological pathways, gene functional classes, etc.) in fitting the model.  We address these two goals by defining a penalty of the form
\begin{equation}
\begin{split}
 P_W(\beta) =
& \lambda_1 \sum_{j=1}^p \left( w_{jj}^2 ||\textbf{X}_j \beta_j||^2 + \sum_{k\ne j} w_{jk}^2 ||\textbf{X}_{jk}\beta_{jk}||^2 \right)^{1/2} \\
& + \lambda_2 \sum_{j=1}^p \sum_{k>j} w_{jk} ||\textbf{X}_{jk}\beta_{jk}|| \enskip ,  \label{eq:02}
\end{split}
\end{equation}
where the $w_{jk}\ge 0$ are non-negative weights provided by the investigator and $W = [w_{jk}]$ is used to denote the matrix of weights over all SNP pairs $i,j$.  The values $\lambda_1, \lambda_2 > 0$ are tuning parameters.

Our penalty is a generalization of that proposed by~\cite{radchenko2010variable} for the purpose of fitting general types of interaction models. (In \cite{radchenko2010variable}, $w_{jk}\equiv 1$ for all $j,k$.) Note that,
following those authors, we express the penalty in un-normalized form.  (Standard lasso algorithms, for example, without interactions, assume $||\mathbf{X}_j||=1$ and hence $||\mathbf{X}_j\beta_j||^2 = \beta_j^2$). It can be shown that the penalty automatically enforces the hierarchical constraint (i.e., inclusion of main effects before interactions). Main effects and interactions can be treated differently by varying $\lambda_2$ with respect to $\lambda_1$. The elements of the matrix $W$ are generic and allow for the possibility of including biological information {\it a priori} into the model selection process. We next describe a manner for doing so, in which network principles are used in a natural way.

\subsection{A Network-Based Penalty}
Below we describe our construction of the matrix $W$ using information from biological pathways, although similar constructions may be obtained quite generally using other common resources (e.g., databases of genes and their biological function, such as Gene Ontology).   Note that $W$ acts as a dissimilarity matrix in $P_W$.  Under our construction, $W$ is defined with respect to a graph showing relationships among SNPs, which in turn derives from a bipartite graph relating SNPs to pathways.  The intuition underlying our construction is (a) to allow interactions only among SNPs corresponding to genes that are common to at least one pathway, and (b) to encourage interactions among those SNP pairs that are common to more pathways.

\begin{figure}[!tpb]
\centerline{\includegraphics[width=8.5cm,height=1.6cm]{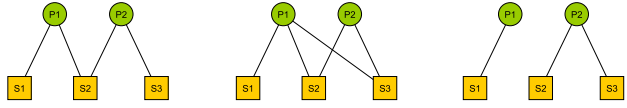}}
\caption{Simple illustration of network representations between SNPs (S1, S2, S3) and pathways (P1, P2).}\label{fig:01}
\end{figure}

\begin{figure}[!tpb]
\centerline{\includegraphics[width=8.5cm,height=2.0cm]{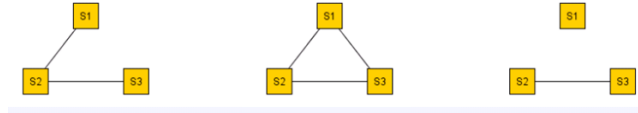}}
\caption{One mode projection of the three examples in Figure~\ref{fig:01}.}\label{fig:02}
\end{figure}

Let $S_1,\ldots, S_p$ denote our $p$ SNPs, and $P_1,\ldots, P_m$, our $m$ pathways.  We define $G$ to be a bipartite graph, with one set of nodes representing SNPs, and the other, pathways.  An edge in $G$ connects a SNP $S_i$ to a pathway $P_{\ell}$ if that SNP maps sufficiently close to a gene found in the pathway.  We then define $G_{SNP}$ to be the one-mode projection of $G$ onto the set of SNPs.  Figures~\ref{fig:01} and~\ref{fig:02} show three toy examples of graphs $G$ and $G_{SNP}$, for $p=3$ SNPs and $m=2$ pathways.

An equivalent representation of the relationship between SNPs and pathways in the network $G_{SNP}$ is a $p\times m$  incidence matrix $M$, describing which SNPs are linked to which pathways. For the three examples in Figure \ref{fig:01}, the corresponding incidence matrices are
\begin{equation}
M_1 = \left[ \begin{array}{cc}
		1 & 0 \\
		1 & 1 \\
		0 & 1 \end{array} \right]
\quad
M_2 = \left[ \begin{array}{cc}
		1 & 0 \\
		1 & 1 \\
		1 & 1 \end{array} \right]
\quad
M_3 = \left[ \begin{array}{cc}
		1 & 0 \\
		0 & 1 \\
		0 & 1 \end{array} \right] \enskip .  \label{eq:03}
\end{equation}
Similarly, the analogous $m\times m$ (weighted) adjacency matrix is the standard representation of the one-mode projection $G_{SNP}$. Calling this matrix $A$, it is related to the incidence matrix $M$ of the original graph $G$ through the expression $A=MM^T$.  For the three examples shown in Figure \ref{fig:01} and Figure \ref{fig:02}, the adjacency matrices are
\begin{equation}
A_1 = \left[ \begin{array}{ccc}
		1 & 1 & 0 \\
		1 & 2 & 1  \\
		0 & 1  & 1 \end{array} \right]
\quad
A_2 = \left[ \begin{array}{ccc}
		1 & 1 & 1 \\
		1 & 2 & 2 \\
		1 & 2 & 2 \end{array} \right]
\quad
A_3 = \left[ \begin{array}{ccc}
		1 & 0 & 0\\
		0 & 1 & 1\\
		0 & 1 & 1 \end{array} \right] \enskip . \label{eq:04}
\end{equation}

Finally, we define the dissimilarity matrix $W$ elementwise by setting $w_{jk}=1/a_{jk}$.  In the case where $a_{jk}=0$, we set $w_{jk}=\infty$ by convention.  Note that the resulting implication for the optimization in (\ref{eq:06}) is that $\beta_{jk}$ is set to zero, i.e., the term $X_{jk}$ cannot enter the model.  Hence, only those pairs of SNPs $j,k$ that share at least one pathway (i.e., $w_{jk}< \infty$) may potentially enter the model.  As a result, it is possible to substantially reduce the number of interaction terms considered for entry into the model, thus making the simultaneous search for main effects and interactions easier to perform.  For example, in the application presented in Section~4, a total of 17,025 SNPs were used, corresponding to nearly 145 million interactions. However, in using the 186 pathways from the KEGG (Kyoto Encyclopeida of Genes and Genomes) database to construct our matrix $W$, this number was reduced to less than 480,000 potential interactions.

We note that there are certainly other ways of constructing the matrix $W$.  For example, a variation on the procedure described above would be to define $w_{jk} = 1$ if $a_{jk}>0$, and infinity otherwise.  This is equivalent to equipping the graph $G_{SNP}$ with a binary adjacency matrix and letting $w_{jk}=1/a_{jk}$ as before, and results in the equal treatment of all interactions that are allowed to enter the model, regardless of how many pathways are shared by pairs $j,k$.  In addition, of course, other types of outside information --- if judged relevant --- can be used in place of pathways, as mentioned above.

\subsection{Model Selection and Fitting}

To perform the optimization in (\ref{eq:06}) we use cyclic coordinate descent, a now-standard choice for problems such as ours (e.g., \cite{wu2008coordinate,friedman2007pathwise,wu2009genome}).  As the name indicates, the cyclic coordinate descent algorithm updates one element of $\mathbb{\beta}$ at a time using coordinate descent principles, while holding all others fixed, and cycles through all elements until convergence.  In our context, the details of the resulting algorithm parallel those of \cite{radchenko2010variable}.  We therefore present only a sketch of the algorithm and relevant formulas here. Detailed derivation can be found in Appendix, Section~A.1.

Consider the estimation of $\beta_j$.  We note that, with respect to this parameter, the objective function in (\ref{eq:06}) can be written as
\begin{equation}
\begin{split}
& \frac{1}{2} \left(\mathbf{\tilde Y}_j - \mathbf{X}_j \beta_j\right)^T
			   \left(\mathbf{\tilde Y}_j - \mathbf{X}_j \beta_j\right) \\
&	+ \lambda_1\left(
	w_{jj}^2 ||\mathbf{X}_j\beta_j||^2 +
	\sum_{k\ne j} w_{jk}^2 ||\mathbf{X}_{jk}\beta_{jk}||^2\right)^{1/2}
	+ \mathcal{C}_j \label{eq:07}
\end{split}
\end{equation}
where $\mathbf{\tilde Y}_j = \mathbf{Y} - \sum_{\ell\ne j}\mathbf{X}_{\ell}\tilde \beta_{\ell} - \sum_{\ell=1}^p \sum_{k>\ell} \mathbf{X}_{\ell k}\tilde \beta_{\ell k}$.  Here $\tilde \beta_{\ell}$ is the current value of $\beta_{\ell}$ at this stage of our iterative algorithm, and similarly for $\tilde \beta_{\ell k}$, while $\mathcal{C}_j$ is all of the rest of the penalty term $P_W(\beta)$ that does not involve $\beta_j$.

The updates to the estimates $\tilde\beta_j$ of the main effects $\beta_j$ take the form of a shrinkage estimate, $\tilde\beta_j=\alpha_j \hat\beta_j$,  for $\alpha_j\in [0,1]$.   Here $\hat\beta_j = \mathbf{X}^T_j \mathbf{\tilde Y}_j$ is the solution to the problem of fitting  a regression-through-the-origin for $\mathbf{\tilde Y}_j$ on $\mathbf{X}_j$, and the shrinkage parameter $\alpha_j$ is the solution to the equation
\begin{equation}
\alpha_j\left(1 + \frac{\lambda_1 w_{jj}^2 }{(w_{jj}^2 \mathbf{X}_j^T \mathbf{X}_j \alpha_j^2\hat\beta_j^2 + c_j)^{1/2}} \right) = 1 \enskip , \label{eq:08}
\end{equation}
where $c_j=\sum_{k\ne j} w_{jk}^2 ||\mathbf{X}_{jk}\beta_{jk}||^2 $.
The value $\alpha_j$ can be obtained using the Newton-Raphson method.  In the special case where $c_j=0$, which must be the case
when $w_{jk}=0$ for all $k\ne j$ (i.e., SNP $j$ is not allowed to participate in any interactions), equation (\ref{eq:08}) can
be solved in closed-form, yielding $\alpha_j = (1- \lambda_1 w_{jj}\, /\, [(\mathbf{X}_j^T \mathbf{X}_j)^{1/2}|\hat\beta_j|])_+ $.

Now consider the estimation of $\beta_{jk}$.  Similar arguments show that the iterations in the cyclic coordinate descent algorithm involve updates of the form $\tilde{\beta}_{jk} = \alpha_{jk}\hat{\beta}_{jk}$, for $\alpha_{jk} \in [0,1]$.  Here $\hat\beta_{jk}=\mathbf{X}_{jk}^T \mathbf{\tilde{Y}}_{jk}\, /\, \mathbf{X}_{jk}^T\mathbf{X}_{jk}$ is the solution to the problem of fitting a regression-through-the-origin for $\mathbf{\tilde Y}_{jk}$ on $\mathbf{X}_{jk}$, where
$$\mathbf{\tilde Y}_{jk} = \mathbf{Y} - \sum_{\ell=1}^{p}\mathbf{X}_{\ell}\tilde{\beta}_{\ell} - \sum_{m>\ell}\sum_{(\ell,m)\neq (j,k)} \mathbf{X}_{\ell m}\tilde {\beta}_{\ell m} \enskip .$$
The shrinkage parameter $\alpha_{jk}$ for interaction terms is the solution to the equation
\begin{equation}
\begin{split}
& \alpha_{jk}\hat{\beta}_{jk}\left\{1+\lambda_1 w_{jk}^2 \left[\frac{1}{(w_{jk}^2\mathbf{X}_{jk}^T\mathbf{X}_{jk}\alpha_{jk}^2\hat{\beta}_{jk}^2+c_1^{jk})^{1/2}} \right.\right.\\
& \left.\left. \qquad +\frac{1}{(w_{kj}^2\mathbf{X}_{kj}^T\mathbf{X}_{kj}\alpha_{jk}^2\hat{\beta}_{jk}^2+c_2^{jk})^{1/2}}\right]\right\} \\
& =sign(\hat{\beta}_{jk})\left[|\hat{\beta}_{jk}|-\lambda_2 w_{jk}(\mathbf{X}_{jk}^T\mathbf{X}_{jk})^{-1/2}\right]_+ \enskip ,
\end{split} \label{eq:09}
\end{equation}
where
$$c_1^{jk}=w_{jj}^2 \mathbf{X}_j^T \mathbf{X}_j \beta_j^2+\sum_{n\neq j,k} w_{jn}^2 \mathbf{X}_{jn}^T\mathbf{X}_{jn}\beta_{jn}^2$$
and
$$c_2^{jk}=w_{kk}^2 \mathbf{X}_k^T \mathbf{X}_k \beta_k^2+\sum_{n\neq k,j}w_{kn}^2\mathbf{X}_{kn}^T\mathbf{X}_{kn}\beta_{kn}^2\enskip ,$$
which again can be computed using the Newton-Raphson method.
When $c_1^{jk}$ and $c_2^{jk}$ are both zero, $\alpha_{jk}$ can be solved in closed form, yielding
$$\alpha_{jk} =\left\{1-[(2 \lambda_1 + \lambda_2) w_{jk}]\, /\, [(\mathbf{X}_{jk}^T \mathbf{X}_{jk})^{1/2} |\hat\beta_{jk}| ]\right\}_+ \enskip .$$

The shrunken estimates of coefficients of predictors and interactions are updated in the iterative process described above until convergence is achieved. Following standard practice, upon termination of our cyclic coordinate descent algorithm we generate a final estimate of coefficients for those variables $X_j$ and $X_{jk}$ that were allowed to enter the model, using ordinary least squares.  All corresponding effect-size estimates and $p$-values produced by our methodology result from this final step.

For datasets with a small number of predictors $\{X_j\}$, the algorithm can be easily fit as described. But for larger numbers of predictors, we employ a `swindle', in analogy to that proposed by~\cite{wu2009genome} and implemented in Mendel (\cite{Lange2001}).  The basic idea is to apply the algorithm to a much smaller number, say $k$, of pre-screened predictors, and to choose the smoothing parameter(s) such that only a desired number, say $s<k$, of predictors $X_j$ enters the model.  The Karush-Kuhn-Tucker (KKT) conditions for our optimization problem are then checked for the estimate $\tilde \beta$ resulting from our algorithm (augmented with zeros for coefficients of all predictors eliminated at the pre-screening stage).  If the KKT conditions are satisfied, we are done; if not, we double $k$ and repeat the process.  Following~\cite{wu2009genome}, we let our initial choice of $k$ be a multiple of $s$, i.e., $k=10\times s$ in the applications we show.  Pre-screening consists of sorting the $t$ statistics of fitting ordinary least-square regression of $Y$ on each predictor $X_j$ separately (i.e., traditional GWAS) and extracting those predictors with the $k$ largest $t$ statistics.  Details can be found in Appendix, Section~A.2.

\subsection{Choice of Tuning Parameters}

In the penalty function $P_W$ defined in (\ref{eq:02}), the tuning parameters $\lambda_1, \lambda_2$ directly influence the number of variables that enter the final model. In principle these two parameters may be allowed to vary freely and a cross-validation strategy used to select the best values.  However, this strategy is unrealistic for GWAS, where the number of SNPs may range into millions.  Instead, we employ a strategy that allows investigators some control in dictating how many variables enter the model, and thereby specify the tuning parameters implicitly.

First, we impose a linear relation between the two tuning parameters, i.e.,  $\lambda_2 = c \lambda_1$. Because $\lambda_2$ is directly involved only in the selection of  interaction terms, specifying the constant $c$ may be interpreted as ``tuning'' the number of interactions relative to main effects. The tuning parameter $\lambda_1$  is responsible for the number of main effects in the model. Since $\lambda_1$ is essentially a decreasing function of the number of main effects entered in the model and often investigators have at least some rough expectation of how many SNPs they feel are likely to be associated with their phenotype, we set $\lambda_1$ by pre-specifying the number of main effects to include in the final model (i.e., denoted $s$ above).

Second, calculations show that the relation $c\approx \frac{\sqrt{\sigma_j \sigma_k}}{r}$ holds, where $\sigma^2_j=2 p_j(1- p_j)$ is the variance of SNP $j$ coded as the number of minor alleles under the assumption of Hardy-Weinberg equilibrium;  the variance is defined here in terms of the minor allele frequency $p_j$, and $r$ is the ratio of the thresholds for main effects and interactions to enter the model within the cyclic coordinate descent algorithm.  See Appendix, Section~A.3, for details.   We recommend that $c$ be chosen by the user through (a) specifying a desired ratio $r$, and (b) knowledge of the distribution of SNP minor allele frequencies.

By setting the desired number of main effects and the value $c$, we implicitly specify the values of the tuning parameters $\lambda_1, \lambda_2$.  A smaller value of $c$ (corresponding to a larger value of $r$) means more interactions may enter the model, for a fixed number of main effects.

\section{Simulations} \label{sec:simulation.main}
\subsection{Simulation Study Design}
We carried out a simulation study in order to assess (i) the performance of our method under various interaction scenarios, and (ii) the effect of different choices of the $W$ matrix in our penalty on our ability to detect interaction.  We also compared our method to the stage-wise selection method proposed by~\cite{wu2009genome}, which restricts interaction search to SNPs first declared to have main effects. In each simulated data set, there are 1000 subjects and 1000 SNPs as predictors. The SNPs are coded additively ($0$,$1$,$2$), simulated with a minor allele frequency (MAF) of $50\%$, and drawn from a Binomial distribution with two trials. Lower MAFs were also investigated (MAF $>=10\%$, see additional simulation in Section~\ref{subsec:simulation.varyMAF}). The quantitative trait $Y$ is then simulated using the effect SNPs and interactions specified under assumed models. Among the 1000 SNPs, 20 (SNP1-SNP20) have true main effects on the simulated trait and the remaining 980 have no effect.

To test our method in various interaction situations, we evaluate three different models:
\begin{itemize}
\item Model 1: only 20 main effects with no interaction
\item Model 2: 20 main effects + all two way interactions among SNP1-SNP5
\item Model 3: 20 main effects + SNP1$\times$SNP2 + SNP3$\times$SNP4 + SNP5$\times$SNP6 + ... + SNP19$\times$SNP20
\end{itemize}
Model 1 has no interactions involved. Models 2 and 3 both have 10 interaction terms involved, and the interactions are all among true main effects. But in Model 2 there is one cluster with 5 interacting SNPs, while in Model 3 there are 10 clusters, each with two interacting SNPs.

In addition, we explore six different ways to construct the $W$ matrix used in the penalty. In each case, we allow all SNPs to be evaluated as possible main effects, by having all ones down the diagonal of $W$.  For the possible interaction terms, coded by the off-diagonal elements of $W$, we consider the following additions
\begin{itemize}
\item $W_1$:  + true interactions in models
\item $W_2$:  + two way interactions among all true main effects (SNP 1-20)
\item $W_3$:  + true interactions + random `noise' interactions
\item $W_4$:  + two way interactions among all true main effects + random `noise' interactions
\item $W_5$:  + two way interactions among SNPs 1-40 (all true main effects and 20 non-active SNPs)
\item $W_6$:  + two way interactions among SNPs 1-10,21-30 + two way interactions among SNPs 11-20,31-40
\end{itemize}

The matrix $W_1$ is an ideal case. It only allows true interactions built in the model to enter that model. Note that $W_1$ is different for each of Models 1, 2, and 3. The matrix $W_2$ introduces some `noise' interactions by allowing all interactions among true main effects.  It is equivalent to a single pathway of SNPs 1-20 and is the same for all models. The matrix $W_3$ adds random `noise' interactions to $W_1$, while $W_4$ adds random `noise' interactions to $W_2$. Note that $W_3$ and $W_4$ both vary across models. The random `noise' interactions are introduced in a manner aimed at  mimicking the interaction structure corresponding to the KEGG (Kyoto Encyclopedia of Genes and Genomes) database, only some subset of which will likely be relevant to any given study (and the rest, `noise').  Specifically, an additional set of `pathways' (i.e., gene sets) were defined, in addition to those defined by the models themselves, until a total of $20$ pathways were formed.  To these $20$ we then randomly allocated $160$ additional SNPs so that the average number of SNPs per pathway roughly mimicked what is observed in KEGG. $W_5$ represents a single pathway of SNPs 1-40, similar to $W_2$ but with more SNPs (20 non-active SNPs) involved. $W_6$ then represents two pathways with each having 10 active and 10 non-active SNPs. It is similar to $W_5$ in the sense that the allowed interactions involve SNPs 1-40, but $W_6$ has smaller amount of non-active interactions.

We chose $\lambda_1$ by setting the desired number of main effects selected as 25, the value of $\lambda_1$ is automatically determined by our program once the value 25 is provided. This is a natural choice since there are 1000 SNPs in our data and 20 true main effects in the models. This choice will affect Type I error because at least 5 of the 25 predictors selected as main effects will be false, but this number is modest compared to the total of 1000 SNPs and can be easily adjusted by re-setting $\lambda_1$ according to the investigator\rq{}s preference. The parameter $c$ is set to 0.5 (i.e., $r=1.0$ under our model). The selected predictors are then ranked by their absolute $t$-values resulting from the ordinary least-square fit on the selected predictors for the final model. By setting a threshold on the rank we choose the number of interactions to be reported and compare the performance of interaction selection under various $W$ matrix specifications across a range of thresholds.

\subsection{Simulation Results} \label{subsec:simulation.result}
\begin{figure*}[!tbp]
\centering
\includegraphics[width=14cm,height=7cm]{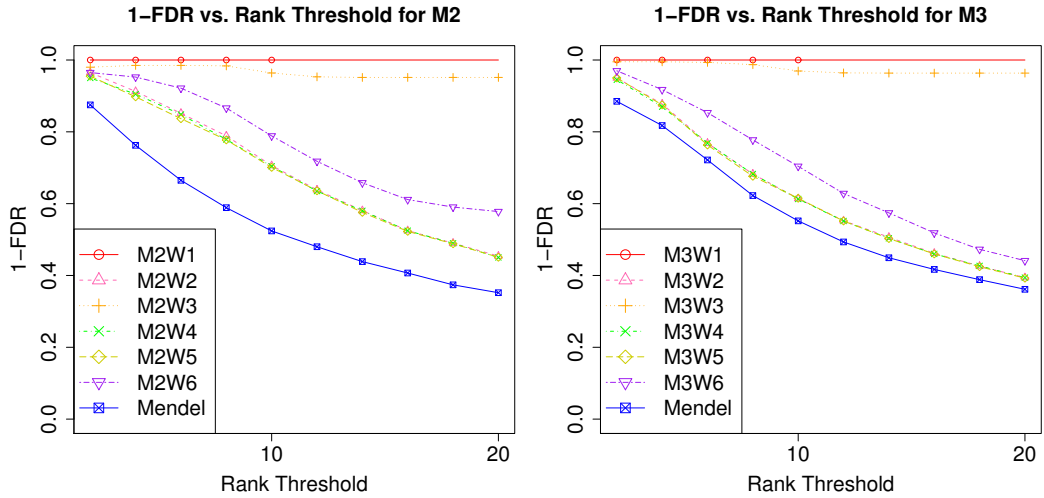}
\caption{Interaction Results for Model 2 and Model 3 with 6 W matrix specifications and Mendel analysis.}\label{fig:03}
\end{figure*}

In Figure \ref{fig:03}, we compare the results under various $W$ matrix specifications, for Models 2 and 3.  We assess the ability to find true interactions by computing the average false discovery rate of interactions over 100 trials and plotting 1-FDR against the rank-threshold for selected interactions. As the threshold increases, more interactions get selected and thus FDR increases and the curves have a downward trend. Examining the results, we see that $W_1$ clearly has the best performance, as it reflect the truth about the interactions in the model; all false interactions are excluded {\it a priori} and thus the 1-FDR curve for $W_1$ is a straight line at 1. Recall that $W_3$ is equivalent to $W_1$ plus random `noise'.  Importantly, therefore, we note that pure `noise' among non-active SNPs does not appear to impact much the selection of true interactions, as $W_3$ has the second best performance after $W_1$.   This conclusion is reinforced by the results for $W_2$ and $W_4$, where the 1-FDR curves are nearly identical.  On the other hand, the results in Figure~\ref{fig:03} also suggest that selection of interactions is to some extent adversely affected when allowing `noise'  interactions among active SNPs,  as $W_6$ has a better performance than $W_2$ and $W_5$ while $W_2$ and $W_5$ have very similar performance.

In comparing our method to that of \cite{wu2009genome}, as implemented in Mendel, we can see in Figure \ref{fig:03} that our method outperforms stage-wise selection for all choices considered for the matrix $W$.  This observation is important in showing that using accurate prior information, even with moderate `noise' (i.e., specifying non-existent interactions), it is possible to out-perform the stage-wise approach by over $10-20\%$ on the 1-FDR scale. Note that we used the default option in Mendel that tests interactions among selected main effects. There are other options in Mendel one can choose that may perform somewhat better.

With respect to the detection of main effects,  the performance of our methodology is shown in Table~\ref{tab:main.effects}.  The average power of main effects are grouped into three categories: the true SNPs involved in interaction, true SNPs not involved in interaction and the SNPs that have no effect on the simulated trait. Recall that there is no interaction in Model 1 and all true SNPs in Model 3 are involved in interaction, so they have only two relevant groups of SNPs. As we can see from the Model 2 result, SNPs involved in interactions are detected more easily than SNPs not involved in interactions. Comparing Table~\ref{tab:main.effects} to Table~\ref{tab:main.effects.mendel}, we can also see that our method has the same or higher  average power to detect true main effects than the stage-wise approach of~\cite{wu2009genome}, as implemented in Mendel. In both approaches the non-active SNPs have a very small  chance of being selected as main effects.

\begin{table}
\centering
{\footnotesize
\begin{tabular}{p{2.5cm} cccccc}
                     & \multicolumn{6}{c}{Model 1} \\  \cline{2-7}
 Main Effects &  $W_1$ &  $W_2$ & $W_3$ & $W_4$ & $W_5$  &  $W_6$  \\ \hline
SNPs w/ Interaction & - & - & - & - & - & -  \\
SNPs w/o Interaction & 0.618 & 0.645 & 0.616 & 0.645 & 0.645 & 0.636 \\
Non-active SNPs  & 0.013 & 0.012 & 0.013 & 0.012 & 0.012 & 0.012 \\ \hline
\end{tabular}

\vspace{0.1in}
\begin{tabular}{p{2.5cm} cccccc}
                     & \multicolumn{6}{c}{Model 2} \\  \cline{2-7}
 Main Effects &  $W_1$ &  $W_2$ & $W_3$ & $W_4$ & $W_5$  &  $W_6$  \\ \hline
SNPs w/ Interaction & 1.000 & 1.000 & 1.000 & 1.000 & 1.000 & 1.000  \\
SNPs w/o Interaction & 0.565 & 0.607 & 0.565 & 0.607 & 0.606 & 0.595 \\
Non-active SNPs  & 0.011 & 0.011 & 0.012 & 0.011 & 0.011 & 0.011 \\ \hline
\end{tabular}

\vspace{0.1in}
\begin{tabular}{p{2.5cm} cccccc}
                     & \multicolumn{6}{c}{Model 3} \\  \cline{2-7}
 Main Effects &  $W_1$ &  $W_2$ & $W_3$ & $W_4$ & $W_5$  &  $W_6$  \\ \hline
SNPs w/ Interaction & 1.000 & 1.000 & 1.000 & 1.000 & 1.000 & 1.000  \\
SNPs w/o Interaction & - & - & - & - & - & - \\
Non-active SNPs  & 0.005 & 0.005 & 0.005 & 0.005 & 0.005 & 0.005 \\ \hline
\end{tabular}
}
\caption{Simulation results for detection of main effects. \label{tab:main.effects}}
\end{table}

\begin{table}
\centering
{\footnotesize
\begin{tabular}{p{2.5cm}cc}
Main Effects &  Model 2  & Model 3 \\ \hline
SNPs w/ Interaction & 1.000  & 1.000 \\
SNPs w/o Interaction & 0.557  &  -  \\
Non-active SNPs & 0.012 & 0.005 \\ \hline
\end{tabular}
}
\caption{Detection of main effects by stage-wise competitor. \label{tab:main.effects.mendel}}
\end{table}

\begin{figure*}[!tbp]
\centering
\includegraphics[width=14cm,height=7cm]{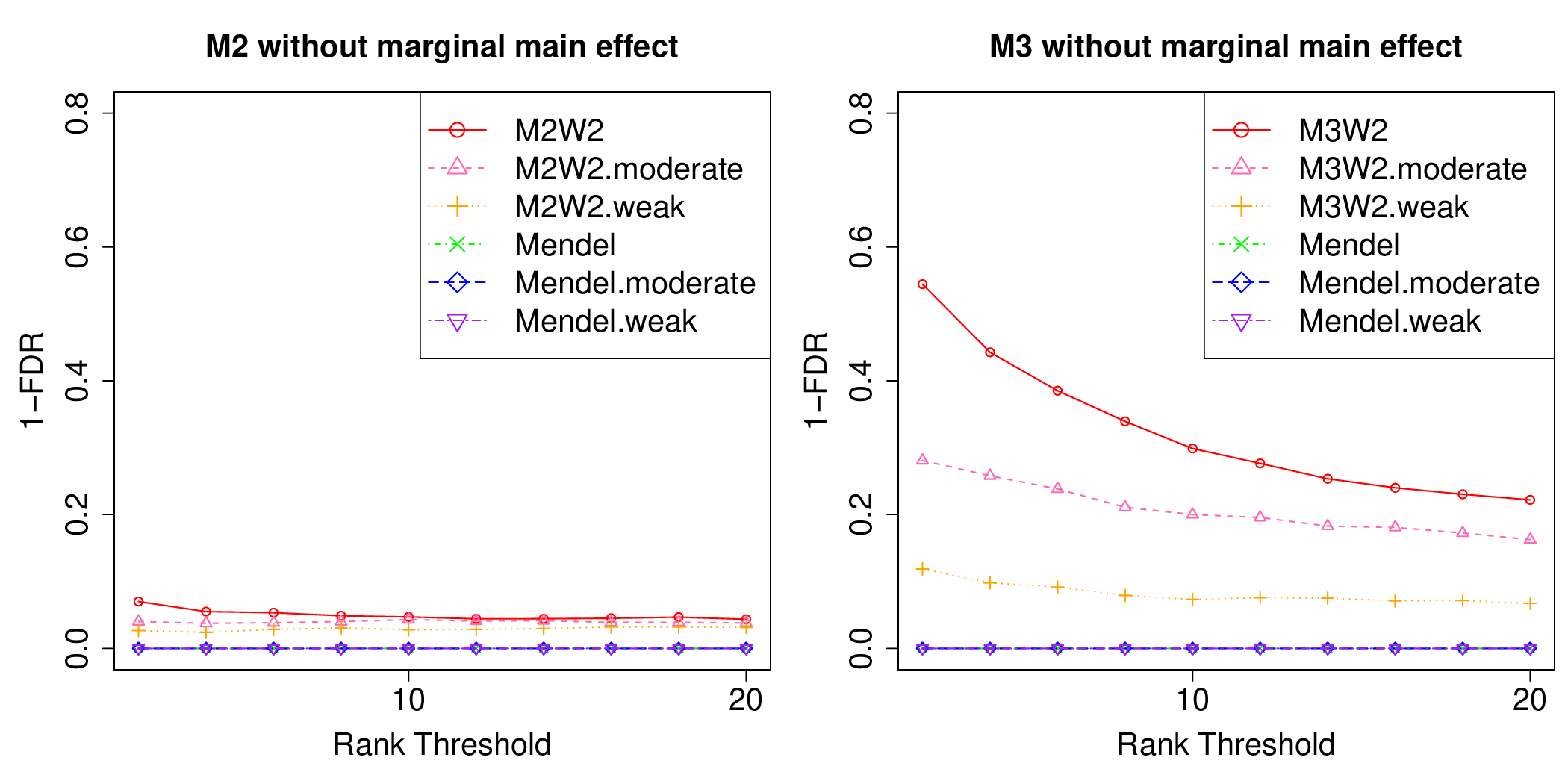}
\caption{Interaction results for Model 2 and Model 3 without marginal main effect.}\label{fig:04}
\end{figure*}

The results just described correspond to simulations of our models where the effect sizes of main effect and interaction were set for $80\%$ power at a type I error rate of $5\%$ under standard single-SNP models with additive-trait structure and SNPs with $50\%$ MAF. Our approach performed similarly with common SNPs with lower MAF (MAF $>=10\%$) and equivalent power, see Section~\ref{subsec:simulation.varyMAF} Figure~\ref{fig:08}. We also test two more cases where main effect sizes were moderate and weak, corresponding to $50\%$ power and $20\%$ power, respectively. To assess the performance of our method in finding interactions under these various strengths of main effects, we reverse the direction of interactions so that there is no marginal SNP effects.  The $W$ matrix we used is $W_2$, as described before, to make a fair comparison with respect to the inclusion of noise interactions. The results under such models are shown in Figure \ref{fig:04}. As we can see from the figure, the approach implemented using Mendel could not find true interactions under any of the models (the regular (Mendel),  the moderate (Mendel.moderate) and the weak (Mendel.weak) main effect models), as it only searches for interactions among main effects selected in the first stage. In contrast, our proposed approach is able to find some of the true interactions because it incorporates information from the $W$ matrix, the network of interactions built from outside knowledge. Not surprisingly, the model with stronger main effect (M2W2, M3W2) performs better in finding true interactions than moderate (M2W2.moderate, M3W2.moderate) or weak (M2W2.weak, M3W2.weak) main effect models.

\section{Additional Simulations} \label{sec:simulation.more}

There are a variety of additional questions that we explored computationally. In this section we present the results of four additional simulations. 

\subsection{Comparison with Simple Association Tests}  

To compare our approach to some simple association tests, we implement two additional methods in simulation. First, we implement a method that tests all main effects using simple linear regression (one SNP at a time) and tests all interactions within the network (i.e., allowed in $W$ matrix) with their main effects. Second, we implement a method that tests all main effects, ranks them based on p-values, and selects the first 25(i.e., the same number as our approach), after which interactions within the selected SNPs are tested, among those that are also allowed by the $W$ matrix. These two approaches and our proposed method are applied to Model 2 with $W_4$.  Tuning parameters for our method are chosen in the same way as before (i.e., setting $\lambda_1$ implicitly be specifying 25 main effects be selected, and setting the parameter $c$ to 0.5).  Results are shown in a 1-FDR plot.

\begin{figure*}[!tbp]
\centering
\includegraphics[width=9cm,height=9cm]{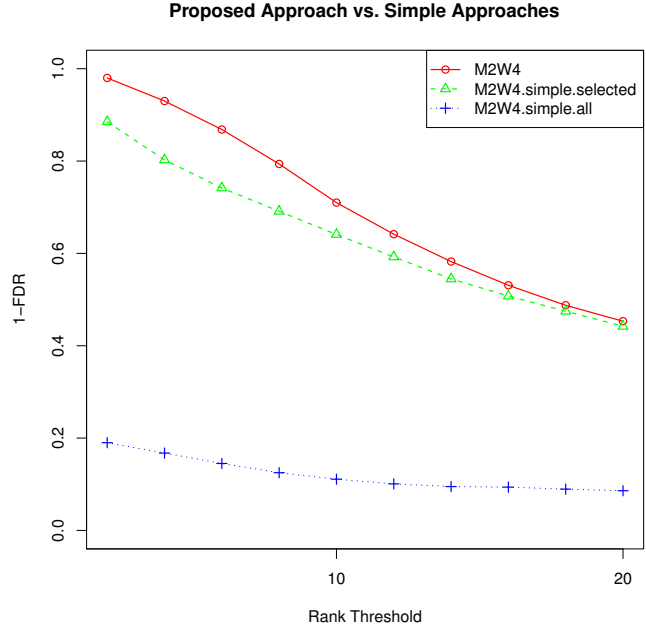}
\caption{Interaction Results for Model 2 with W4 for our proposed method vs. simple association tests.}\label{fig:05}  
\end{figure*}

\begin{table}
\centering
{\footnotesize
\begin{tabular}{p{3cm}ccc}
Main Effects &  Our approach  & Simple test 1 & Simple test 2 \\ \hline
SNPs w/ Interaction & 1.000  & 1.000 & 1.000\\
SNPs w/o Interaction & 0.607  &  0.251  & 0.251\\
Non-active SNPs & 0.011 & 0.017 & 0.017\\ \hline
\end{tabular}
}
\caption{Detection of main effects: Our approach vs. two simple association tests. \label{tab:simple}}
\end{table}

As we can see from Figure \ref{fig:05}, the second method performs better than the first one in finding true interactions. However, it is also very clear that our proposed approach outperforms both of the methods in finding true interactions and main effects (Table \ref{tab:simple}).

\subsection{Stability of Detection with Larger Numbers of SNPs}   

Because of the multitude of conditions we explored through simulations in Section \ref{sec:simulation.main}, and for reasons of computational expediency, we chose to use $p=1000$ SNPs in our models.  However, in practice, substantially larger numbers of SNPs will be used.  One relevant question to examine is whether the detection levels found through simulations with 1000 SNPs remain stable for larger numbers of SNPs.

In order to explore this question, we perform additional simulations for 10,000 SNPs, where 9000 additional 'noise' SNPs are added in the simulated data. The simulation is conducted for Model 2, with $W_4$. The tuning parameters are chosen in the same way as before (i.e., setting $\lambda_1$ implicitly by specifing 25 main effectsbe selected, and setting the parameter $c$ to 0.5).

\begin{figure*}[!tbp]
\centering
\includegraphics[width=9cm,height=9cm]{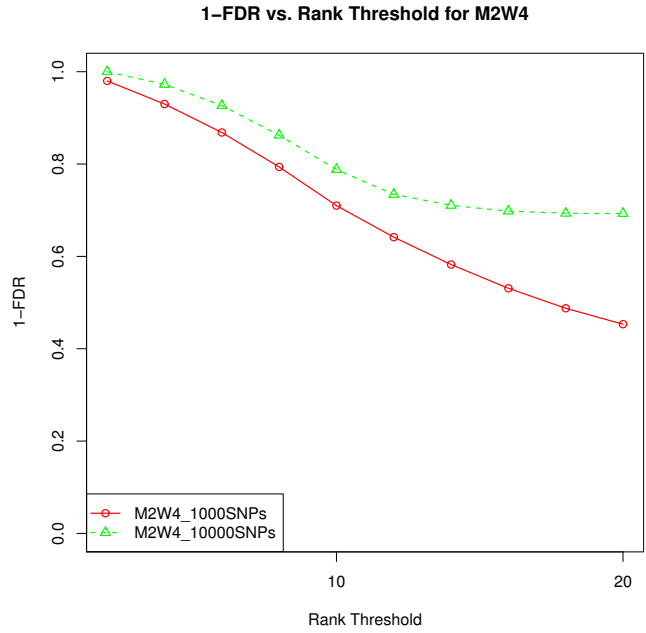}
\caption{Comparison of Interaction Results for Model 2 with $W_4$: 1000SNPs vs. 10000SNPs .}\label{fig:06}  
\end{figure*}

\begin{table}
\centering
{\footnotesize
\begin{tabular}{p{3cm}cc}
Main Effects &  Analysis with 1000 SNPs  & Analysis with 10,000 SNPs \\ \hline
SNPs w/ Interaction & 1.000  & 1.000 \\
SNPs w/o Interaction & 0.607  & 0.309   \\
Non-active SNPs & 0.011 &  0.002 \\ \hline
\end{tabular}
}
\caption{Detection of main effects: Analysis with 1000 SNPs vs. Anslysis with 10,000 SNPs. \label{tab:scaleup}}
\end{table}

Our results show that, as expected, with a greater number of  'noise' SNPs, it is harder to find true main effects (lower percentage of finding true main effects, shown in Table \ref{tab:scaleup}).  However, interestingly, our results also show that, if anything, there is a slightly {\em higher} rate of true discoveries among declared interactions.  This perhaps surprising result can be explained as follows.  As we mentioned in Section~\ref{subsec:simulation.result}, pure 'noise' interactions among 'noise' SNPs does not have a large  impact on the selection of true interactions, but allowing 'noise' interactions among active SNPs adversely affects the selection of true interactions.  At the same time, the number of true interactions selected remained roughly constant in scaling from 1000 to 10,000 SNPs.  Hence the rate of true discoveries among interactions is slightly higher.  See Figure~\ref{fig:06}.

\subsection{The Relative Importance of Network Information}

Note that the penalty used in our method accomplishes two goals simultaneously: it enforces a hierarchical constraint on the inclusion of terms in the model (i.e., interactions after main effects), and it uses network information to restrict which interactions are considered.  To evaluate the relative importance of enforcing hierarchical structure versus incorporating network information, we performed a simulation with $W=II^T$, ($I=(1,1,\ldots,1)^T$) -- where the network information is ignored and all interactions are allowed, compared to a $W$ matrix incorporated some network information (i.e., on top of enforcing hierarchical structure). Because the assessment of interactions among all SNPs is computationally burdensome, we perform this simulation on a reduced version of the model.  Specifically, we simulate 100 SNPs with 10 true main effects (SNP1-10). The interaction structure is the same as Model 2 (all two way interactions among SNP1-5).

Three analyses are compared:
\begin{enumerate}
\item  $W_{all}=II^T$:  network information ignored, all possible interactions are allowed
\item  $W_{first10}$: two way interactions among all true main effects (SNP1-10), similar to $W_2$ when we had 20 true main effects
\item Mendel : step-wise approach that test interactions among selected main effects (default option in Mendel)
\end{enumerate}
The last method (i.e., Mendel) is the same as described before, and is included here simply for comparison.

\begin{figure*}[!tbp]
\centering
\includegraphics[width=9cm,height=9cm]{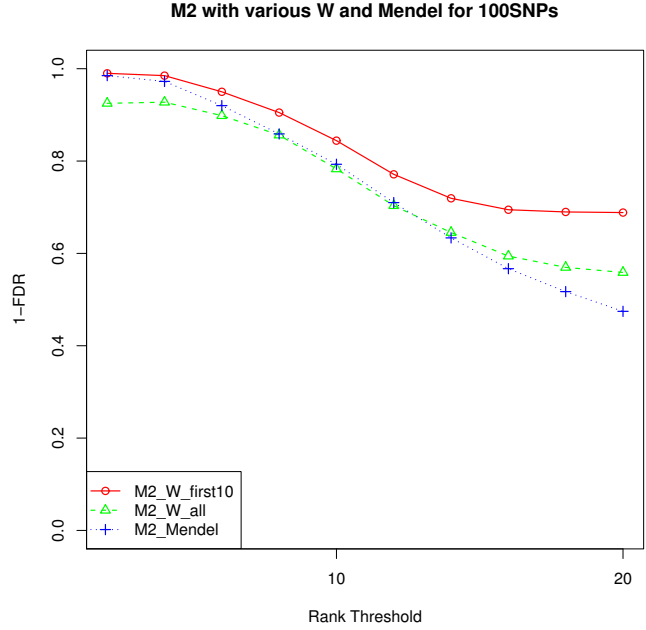}
\caption{Interaction Results for Model 2 under 2 $W$ matrix specifications and Mendel analysis.}\label{fig:07}
\end{figure*} 

Comparing the result with network information ignored (hierarchical feature retained, M2 W.all in Figure \ref{fig:07}) and the result with network information incorporated (M2 W.first10 in Figure \ref{fig:07}), the 1-FDR curve shows that the latter has a better performance in finding true interactions. Also by allowing all interactions, the computing time is dramatically increased. The approach implemented in Mendel (default option) performs better than the analysis with no network information, in finding true interactions when the rank threshold is small (the first few selected interactions have higher percentage of being true), and is also close to the performance of the analysis with network information. When the rank threshold increases (looking at selected interactions further down the list), Mendel performs worse than the analysis with network information and closer to the results without network information. This phenomenon makes sense because the approach in Mendel also has a hierarchical property (searching for interactions among selected main effects).

\subsection{Performance with Varying Minor Allele Frequencies} \label{subsec:simulation.varyMAF}

\begin{table}
\centering
{\footnotesize
\begin{tabular}{p{3cm} ccc|ccc}
                     & \multicolumn{3}{c}{Model 2} &  \multicolumn{3}{c}{Model 3}\\  \cline{2-7}
 Main Effects &  $W_2$ & $W_3$ & $W_4$ &  $W_2$ & $W_3$ & $W_4$  \\ \hline
SNPs w/ Interaction & 1.000 & 1.000 & 1.000 & 0.993 & 0.993 & 0.992  \\
SNPs w/o Interaction & 0.630 & 0.587 & 0.629 & - & - & - \\
Non-active SNPs  & 0.011 & 0.011 & 0.011 & 0.005 & 0.005 & 0.005 \\ \hline
\end{tabular}
}
\caption{Detection of main effects: Analysis of SNPs with varying MAF. \label{tab:varyingMAF}}
\end{table}

\begin{figure*}[!tbp]
\centering
\includegraphics[width=15cm,height=8cm]{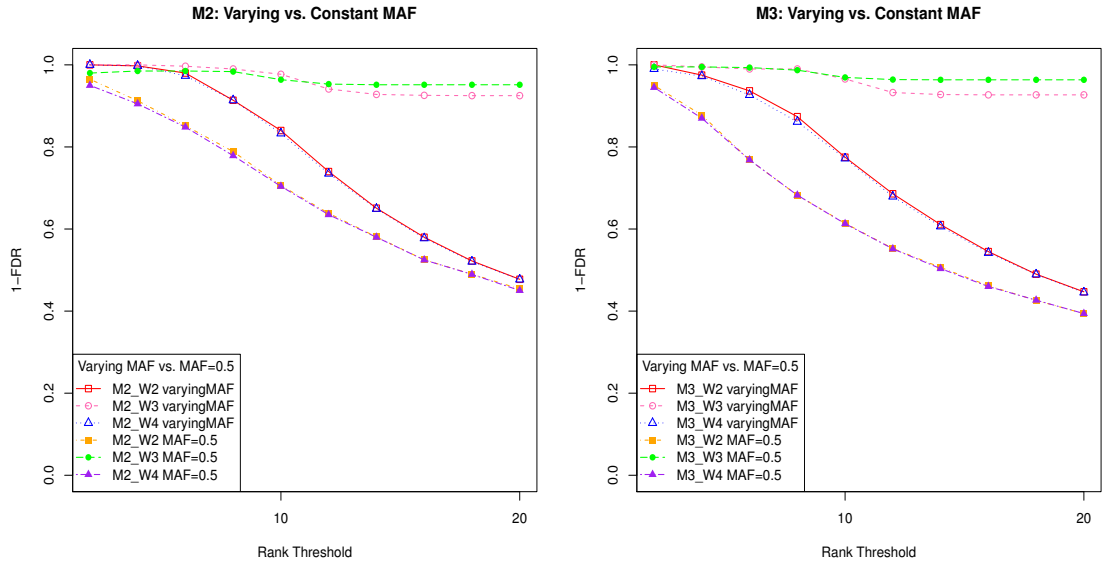}
\caption{Interaction Results for Model 2 and Model 3 with 3 $W$ matrix specifications: Varying MAF vs. Constant MAF.}\label{fig:08} 
\end{figure*}

Although the SNPs were simulated with minor allele frequency (MAF) of $50\%$, we perform additional simulation to explore the robustness of our approach when MAF varies. The MAFs of $10\%$, $20\%$, $30\%$, $40\%$ and $50\%$ are randomly assigned to non-active SNPs. They are also each assigned to 4 of 20 true SNPs (MAF=$10\%$ for SNP1, 6, 11, 16; MAF=$20\%$ for SNP2, 7, 12, 17; MAF=$30\%$ for SNP3, 8, 13, 18; MAF=$40\%$ for SNP4, 9, 14, 19; and MAF=$50\%$ for SNP5, 10, 15, 20). SNPs are coded additively and simulated under a Binomial distribution with two trials, as before.  Keeping the effect sizes of main effects and interactions corresponding to the same power ($80\%$) as before, we tested Model 2 and Model 3 with $W$ matrices $W_2$, $W_3$ and $W_4$ and compared to the result of analysis with MAF of $50\%$, in Figure \ref{fig:08}. As we can see from Figure \ref{fig:08} (for interaction result) and Table \ref{tab:varyingMAF} (for main effect result), the analysis of SNPs with varying MAF has a similar performance compared to the analysis with MAF=$50\%$

\section{Discussion}
There are many potential sources of missing hereditability.  Gene-by-gene interactions is one potential source. In turn, there are many types of genetic interactions, including multiplicative and non-multiplicative \cite{Chatterjee2008,Chatterjee2012}. In this chapter, we focus on investigating multiplicative interactions in the form of a product between two variables.  Our proposed methodology provides a promising new approach to identify such interactions, by exploiting the wealth of biological knowledge accumulated in various pathway databases.

The simulations reported in Section~\ref{sec:simulation.main} suggest that our approach performs better in finding true interactions with a reasonable prior biological knowledge incorporated, compared to the stage-wise regression method that first fits a main effect model and then searches for interactions among selected main effects. The ability of finding true main effects is retained, as compared to the stage-wise approach.

Furthermore, the additional simulations reported in Section~\ref{sec:simulation.more} show that (1) our approach outperforms simple association tests; (2) scaling up data size by adding more `noise' SNPs makes it harder to find true main effects but does not adversely affect the selection of interactions; (3) using network information in our penalty results in decreasing computing time, and also yields advantages in detecting interactions beyond the advantage derived from the hierarchical nature of the penalty; and (4) our approach with varying MAF has a similar performance to the one with constant MAF. 

We implemented our proposed method in R and the code is available at 
\\ {\em http://math.bu.edu/people/kolaczyk/software}

\cleardoublepage

\chapter{Extension to Multiple Cohorts: Meta-Analysis Approaches}
\label{chapter: Extension to Multiple Cohorts: Meta-Analysis Approaches}
\thispagestyle{myheadings}

\section{Motivation}
Meta-analysis is a general approach to combine results from multiple cohorts and is routinely used in Genome Wide Association Studies (GWAS). To increase power to detect true signals, multiple studies are combined to increase sample size. Individual level data usually cannot be pooled among studies because of restrictions due to subjects\rq{} confidentiality, so meta-analysis approaches are often used to combine summary statistics across studies. Widely used meta-analysis approaches include Fisher\rq{}s method \cite{Fisher1925,Fisher1948}, Stouffer\rq{}s Z-score method \cite{Stouffer1949} and inverse variance method using fixed effect model \cite{Hartung2008,Willer2010}. These methods require valid p-values or $\beta$ and SE estimates of participating studies. 

Penalized regression is an effective multivariate approach to select important predictors in GWAS. However, it doesn\rq{}t provide p-values, because the $\beta$ estimates are shrunk due to the penalty involved in optimization and they are used for variable selection (depending on if the term being considered has a non-zero coefficient estimate or not) instead of effect size estimation \cite{Wasserman2009,Meinshausen2009}.  

To extend our proposed methodology to multi-cohort setting, we need to tackle two issues. The first is to obtain valid p-values. Ordinary least squares are often used to obtain p-values or effect size estimation ($\beta$). But regular regression with selected terms cannot be applied to the same set of data that has been used for variable selection. Recent articles have been focused on data-splitting method to obtain p-values for high-dimensional data. \cite{Wasserman2009} proposed splitting the observations into two subsets, using one subset for variable selection and using the other subset to obtain valid p-values. \cite{Meinshausen2009} further suggested multiple splits instead of single split, because the result of single split highly depends on the arbitrary split. We employ data splitting method to obtain p-values for our method proposed in Chapter 2. 

The second challenge is to develop an appropriate procedure to meta-analyze across the cohorts. We propose two extensions to the splitting method for meta-analysis: splitting within cohort and splitting cohorts. The first approach involves splitting data for each cohort, selection of variables on one subset and computation of p-values on the other subset, and finally a meta-analysis across all cohorts. This is a natural extension of our method and splitting method to multiple cohorts. The second approach involves splitting cohorts instead. Cohorts are split into two groups, one group used for variable selection and the other used for obtaining p-values and meta-analysis. This is a more practical approach because it simplifies the communication process among different studies and reduces the possibility of making errors. 

\section{Methods: Pooling Data Across Studies}
Based on the discussion above, we propose to examine four procedures for combining data across multiple cohorts. The variable selection method used should be the same for all of the following four procedures. In our current work, we used the method we proposed in Chapter 2. But the procedures we proposed here for meta-analysis are applicable, in principle, to general penalized regression methods of the form
\begin{equation}
Y = \beta_0 + \sum_{j=1}^p \beta_j X_j  + \epsilon \enskip \label{eq:ch3.01}
\end{equation}
where $\mathbb{\beta}$ is estimated through a penalized least-squares criterion. 
\begin{equation}
\mathbb{\tilde{\beta}} = \arg\min_{\mathbb{\beta}}
	\frac{1}{2}\big|\big| \textbf{Y} - \sum_{j=1}^p \beta_j \textbf{X}_j \big|\big|^2
+ P_W(\mathbb{\beta})\enskip .    \label{eq:ch3.02}
\end{equation}
where $P_W$ is a function of coefficient $\mathbb{\beta}$. And the theoretical justification for the case of penalized regression with the classical Lasso has already been provided \cite{Wasserman2009,Meinshausen2009}.

Again similar to Chapter 2, in model (\ref{eq:ch3.01}) the variables $X_j$ are SNPs (coded as number of minor alleles 0, 1, 2). The goal of the variable selection method is to identify a small set of SNPs that explain the dependent variable $Y$. After variable selection, we estimate p-values (or equivalently $\beta$ and SE) using data-splitting method (each procedure has a different way of splitting data in multi-cohort setting). 

As mentioned earlier, there are several different ways of conducting meta-analysis in a regular regression framework. Fisher\rq{}s method \cite{Fisher1925,Fisher1948} provides a way of combining p-values across studies, but it doesn\rq{}t take into account the direction of the effects. Stouffer\rq{}s Z-score method \cite{Stouffer1949} solved this problem by using Z-scores instead of p-values. And this is the reason that in the following four procedures we obtain Z-scores instead of p-values. 

Unlike Fisher\rq{} method and Stouffer\rq{}s Z-score method, the inverse variance based method \cite{Hartung2008,Willer2010} estimate $\beta$ coefficient (effect sizes) and SE in addition to Z-scores (or equivalently p-values). But not all of the procedures can utilize this method depending on their different data-splitting scheme. So we obtain $\beta$ and SE when available (Z-scores otherwise) out of data-splitting and regular regression steps, and use these as the input for meta-analysis (inverse variance based method when $\beta$ and SE are available, Stouffer\rq{}s Z-score method otherwise). And the result of the meta-analysis will be a set of $\beta$ and SE (or Z-scores) for selected terms in the variable selection step ($\beta=0$ or $Z=0$  for terms not selected). You may also choose to obtain p-values from the final result of the meta-analysis. But in our work, we choose to show the result in Z-scores because it provides effect direction. Different variations on our data-splitting principle are described in the following four procedures.

\subsection{Procedure A: Ideal Situation, the Mega-Analysis}
When analyzing data from multiple studies involved in a consortium, the ideal strategy would be to pool individual data together and conduct analysis as one dataset. Although not practical, it is a \lq{}gold standard\rq{} when comparing other procedures. And our goal is to find the meta-analysis procedure that behaves most similarly to the mega-analysis. We use the following algorithm to conduct Procedure A, the mega-analysis.
  
\begin{enumerate}
\item  For $k=1,2, \ldots, K$, where $K$ is the number of splits,
    \begin{enumerate}
    \item Randomly split the pooled data into two parts $D_1^{(k)}$ and $D_2^{(k)}$ of equal size.
    \item 
         \begin{enumerate} 
          \item Run our selection method using only $D_1^{(k)}$.
          \item Select $s$ SNPs and interactions.
          \end{enumerate}
    \item 
          \begin{enumerate}
          \item Using only $D_2^{(k)}$, fit linear regression with the selected predictors (main effects and interactions) from $D_1^{(k)}$.
          \item Obtain p-values for selected predictors.
          \item Set p-values to 1 for unselected predictors.
          \end{enumerate} 
    \item 
           \begin{enumerate}
           \item Adjust p-values using Bonferroni correction (divide the original p-values by the number of selected predictors).
            \item Convert adjusted p-values to corresponding $Z$ values using standard normal distribution.
            \end{enumerate}
    \end{enumerate}
\item  Average $Z$ scores over $K$ sets of results.
\end{enumerate}

\subsection{Procedure B: Split Cohorts into Two Groups}
A more practical extension of the splitting method is to split the cohorts into two groups. We select variables using cohorts in the first group, then calculate p-values/$Z$ scores and conduct meta-analysis using cohorts in the second group. This is a two-step procedure. In the first step, all cohorts run our method to select variables and report back their selected predictors. Then the meta-analysis center randomize the cohorts into two groups $K$ times (let $K$ be the number of splits) and create $K$ union lists of selected predictors using results reported by cohorts in the first group. In the second step, cohorts are asked to fit final models at most $K$ times, according to the number of times them being assigned to the second group.

This approach makes it easier for individual studies to perform the necessary analyses, and simplify the overall communication process among studies. The algorithm for Procedure B is described below.

\begin{enumerate}
\item  For all cohorts $m= 1,2,\ldots, M$, run our method to select predictors.
\item  For $k=1,2, \ldots, K$,
    \begin{enumerate}
   \item Randomly split cohorts into two equal groups $\{set_{1}^{(k)}\}$ and $\{set_{2}^{(k)}\}$ (each set contains equal number of cohorts)
   \item Use the selected SNPs and interactions from $\{set_{1}^{(k)}\}$ to create a common list (union) of predictors for the current split.
   \item 
         \begin{enumerate} 
         \item Run linear regression for the common list of predictors on $\{set_{2}^{(k)}\}$ cohorts.
         \item Obtain $\beta$ coefficients and SE for predictors on the union list.
         \end{enumerate}
    \item  
          \begin{enumerate}
          \item Conduct meta-analysis using $\beta$ and SE across $\{set_{2}^{(k)}\}$ cohorts.
          \item Calculate $Z=\beta / SE$ for predictors on the union list and let $Z=0$ for unselected predictors.
          \end{enumerate}
    \end{enumerate}
\item   
        \begin{enumerate}
        \item Average $Z$ scores over $K$ splits.
         \item Calculate p-values assuming $Z$ follows standard normal distribution.
         \item Adjust p-values using Bonferroni correction and convert p-values back to $Z$ values.
         \end{enumerate}
\end{enumerate}

\subsection{Procedure C: Split Each Cohort into Two Parts}
The data splitting method \cite{Wasserman2009,Meinshausen2009} suggests splitting the data into two equal subsets, selecting variables using one subset and obtaining valid p-value using the other subset. A natural extension of this method to multiple-cohorts setting is to split data for each cohort and perform meta-analysis using results from each cohort. The following algorithm describes this approach.

\begin{enumerate}
\item  For cohort $m$, $m=1,\ldots,M$,
    \begin{enumerate}
    \item  For $k=1,2, \ldots, K$,  where $K$ is the number of splits,
       \begin{enumerate}
        \item Randomly split the data of cohort $m$ into two parts $D_{m1}^{(k)}$ and $D_{m2}^{(k)}$ of equal size.
        \item 
              \begin{enumerate}
              \item Run our selection method using only $D_{m1}^{(k)}$.
              \item Select $s$ SNPs and interactions.
              \end{enumerate}
        \item 
               \begin{enumerate} 
               \item Using only $D_{m2}^{(k)}$, fit linear regression with selected predictors (main effects and interactions) from $D_{m1}^{(k)}$.
               \item Obtain p-values for selected predictors.
               \item Set p-values to 1 for unselected predictors.
               \item Convert p-values to corresponding $Z$ values in standard normal distribution.
               \end{enumerate}
        \end{enumerate}
    \item  Average $Z$ scores for cohort $m$ over $K$ sets of results.
    \end{enumerate}
\item Conduct meta-analysis using $Z$ scores across $M$ cohorts.
\item 
       \begin{enumerate} 
        \item Adjust for multiple testing. 
        \item  Convert the $Z$ scores (result from meta-analysis) to p-values, adjust p-values using Bonferroni correction, and convert them back to $Z$ scores.
        \end{enumerate}
\end{enumerate}

\subsection{Procedure D: A Variation of Procedure C}
This approach is a variation of Procedure C. In procedure C, each study conducts analysis (variable selection and p-values computation) separately without shared information. The information is aggregated in the last step, the meta-analysis. It is possible and highly likely that the selected predictors and interactions are different among studies. Non-selected predictors are assigned zero $Z$ scores (or equivalently, p-value of $1$). While in Procedure B, there is shared information before obtaining p-values. The list of selected terms is the union of all selected variables in the first group.  To see if the sharing of information makes a difference, we modify Procedure C so that there is information sharing before obtaining p-values. We name the modified version Procedure D.

As we will see in the following description, after variable selection is completed by each cohort on the first half of their data, a union list of all selected terms is created and shared among all cohorts. All the cohorts obtain p-values using the same selected terms on their second half data.

This procedure is even less practical than Procedure C and increases the challenge in the communication process among studies. The main purpose of examining this procedure is to find out if the information sharing makes Procedure C perform differently from Procedure A and B. The algorithm for Procedure D is described below.

\begin{enumerate}
\item  For $k=1,2, \ldots, K$,
    \begin{enumerate}
    \item  For cohort $m$, $m=1,\ldots,M$,
       \begin{enumerate}
        \item Randomly split the data of cohort $m$ into two parts $D_{m1}^{(k)}$ and $D_{m2}^{(k)}$ of equal size.
        \item 
              \begin{enumerate}
              \item Run our selection method using only $D_{m1}^{(k)}$.
              \item Select $s$ SNPs and interactions.
              \end{enumerate}
        \end{enumerate}
    \item Obtain union set of selected terms based on $D_{m1}$ for all $m$.
    \item For cohort $m$, $m=1,\ldots,M$,
        \begin{enumerate}        
        \item Using only $D_{m2}^{(k)}$, regress on the union set of selected terms.
        \item Obtain $\beta$ and SE for predictors in the union set.
        \end{enumerate}
    \item  
            \begin{enumerate}
            \item Conduct meta-analysis for $\beta$ and SE across $M$ cohorts.
            \item Calculate $Z=\beta / SE$ for predictors in the union set and let $Z=0$ for unselected predictors.
            \end{enumerate}
    \end{enumerate}
\item 
        \begin{enumerate} 
        \item Average $Z$ scores over $K$ splits.
        \item Calculate p-values assuming $Z$ follows standard normal distribution.
        \item Adjust p-values using Bonferroni correction and convert p-values back to $Z$ values.
        \end{enumerate}
\end{enumerate}

\section{Simulation}
\subsection{Simulation Study Design}
We conduct a simulation study to compare the performance of all four procedures. Procedure A is the ideal case (mega-analysis), which is what would happen if the datasets from all cohorts could be pooled together and analyzed as one dataset. So our goal is to find the procedure that has similar performance to procedure A. 

There are $M=10$ cohorts in our multi-cohort simulation. We used a similar data generating process as we used in Chapter 2, which we outline below.

Each dataset (cohort) has 1000 subjects and 1000 SNPs as predictors. The SNPs are coded additively ($0$,$1$,$2$), simulated with a minor allele frequency (MAF) of $50\%$, and drawn from a Binomial distribution with two trials. The quantitative trait $Y$ is then simulated using the effect SNPs and interactions specified under the assumed models. Among the 1000 SNPs, 20 (SNP1-SNP20) have true main effects on the simulated trait and the remaining 980 have no effect. The models and $W$ matrices are the same as in Chapter 2.

We evaluate two models:
\begin{itemize}
\item Model 2: 20 main effects + all two way interactions among SNP1-SNP5
\item Model 3: 20 main effects + SNP1$\times$SNP2 + SNP3$\times$SNP4 + SNP5$\times$SNP6 + ... + SNP19$\times$SNP20
\end{itemize}

and five different ways to construct the $W$ matrix used in the penalty:
all SNPs as possible main effects +
\begin{itemize}
\item $W_2$:  + two way interactions among all true main effects (SNP 1-20)
\item $W_3$:  + true interactions + random `noise' interactions
\item $W_4$:  + two way interactions among all true main effects + random `noise' interactions
\item $W_5$:  + two way interactions among SNPs 1-40 (all true main effects and 20 non-active SNPs)
\item $W_6$:  + two way interactions among SNPs 1-10,21-30 + two way interactions among SNPs 11-20,31-40
\end{itemize}

\subsection{Simulation Result}
\begin{figure*}[!tbp]
\centering
\includegraphics[width=15cm,height=18cm]{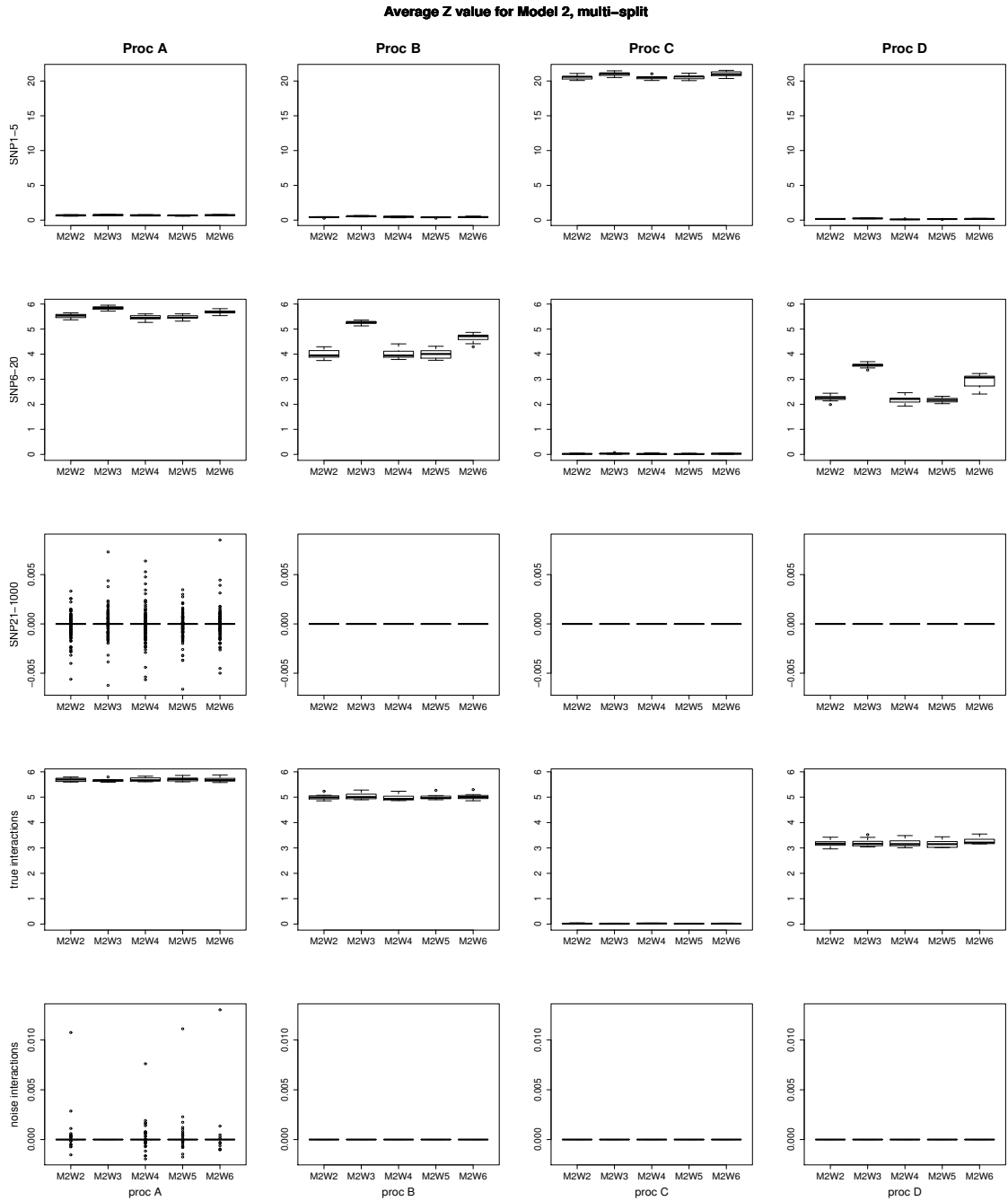}
\caption{Multi-split Model 2: Average Z values for Procedure A, B, C, D with 5 W matrices. Predictors/interactions are categorized into 5 groups: active SNPs involved in interactions, active SNPs not involved in interactions, non-active SNPs, true interactions, noise interactions}\label{fig:3-1}
\end{figure*}

\begin{figure*}[!tbp]
\centering
\includegraphics[width=15cm,height=15cm]{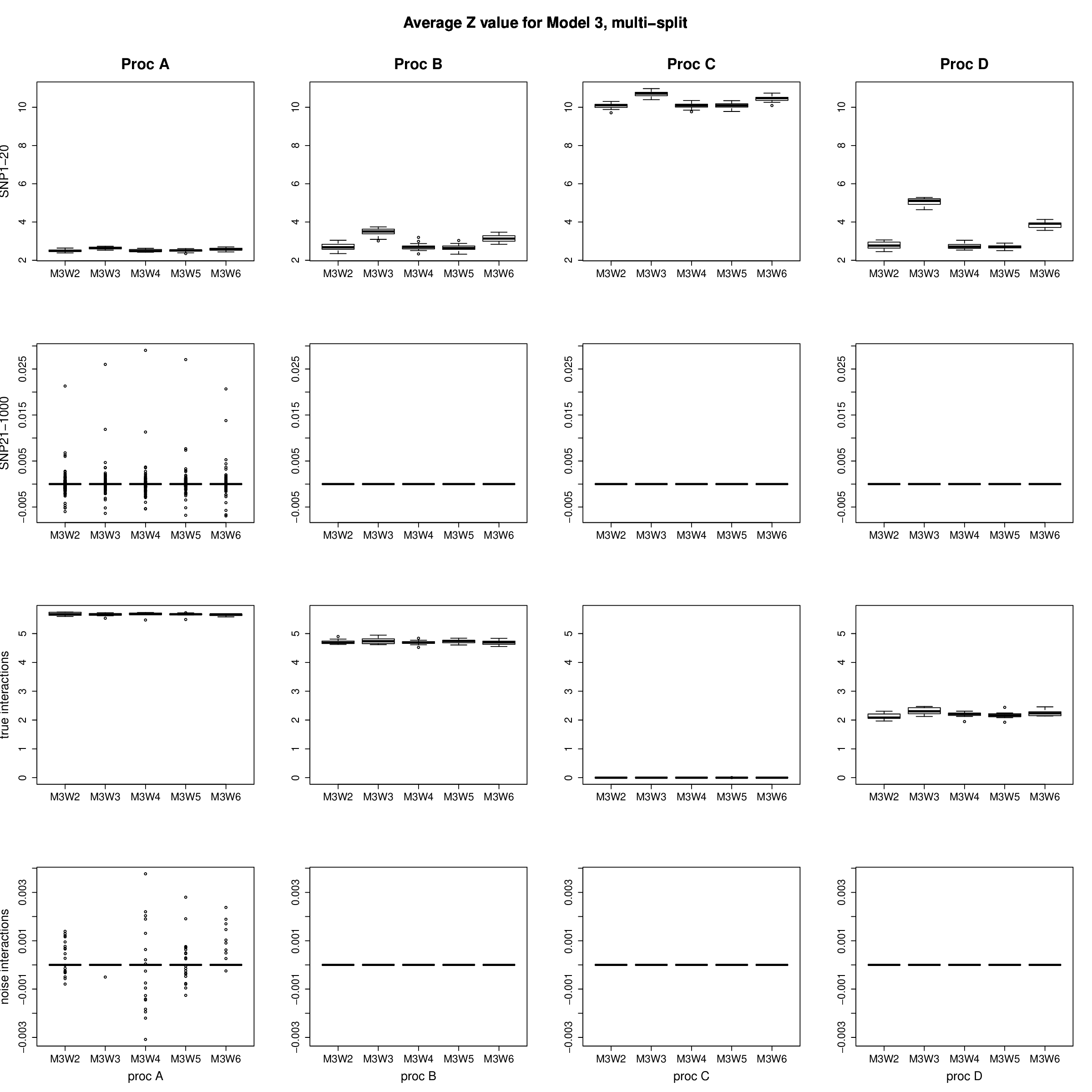}
\caption{Multi-split Model 3: Average Z values for Procedure A, B, C, D with 5 W matrices. Predictors/interactions are categorized into 4 groups: active SNPs, non-active SNPs, true interactions, noise interactions}\label{fig:3-2}
\end{figure*}

The simulation results are summarized in terms of Z scores for all predictors and interactions averaging over 100 simulations. The matrix plots have four columns, each representing one procedure A, B, C, D, in that order. Within each plot, the results under 5 W matrix specifications are shown. Each row of the matrix plots represent one group of predictors/interactions. As shown in Figure \ref{fig:3-1}, predictor and interactions are categorized into 5 groups for Model 2:  active SNPs involved in interactions, active SNPs not involved in interactions, non-active SNPs, true interactions, noise interactions. In Figure \ref{fig:3-2} of Model 3 , predictors and interactions are categorized into 4 groups: active SNPs, non-active SNPs, true interactions, noise interactions, because all active SNPs are involved in true interactions. 

Comparing results from all procedures for each group of predictors/interactions, Procedure B is the one that performs most closely to Procedure A. There is an obviously difference in patterns between Procedures B and C, when comparing them to Procedure A. When the true main effects are also involved in true interactions, Procedure C tends to select main effects rather than interactions. In Figure \ref{fig:3-1}, Procedure C has very high Z values for SNP1-5 while very low Z values for true interactions, which is exactly the opposite to the performance of Procedure A and B. Procedure C also has low Z values for active SNPs 6-20 not involved in interactions. By modifying Procedure C to share information before obtaining valid p-values (i.e. Procedure D), the performance of Procedure D is much closer to that of Procedure A, although still not as close as Procedure B. This is an interesting phenomenon since Procedure B is also the most practical procedure. 

The same conclusion holds for Model 3, in Figure (\ref{fig:3-2}).

\begin{figure*}[!tbp]
\centering
\includegraphics[width=15cm,height=18cm]{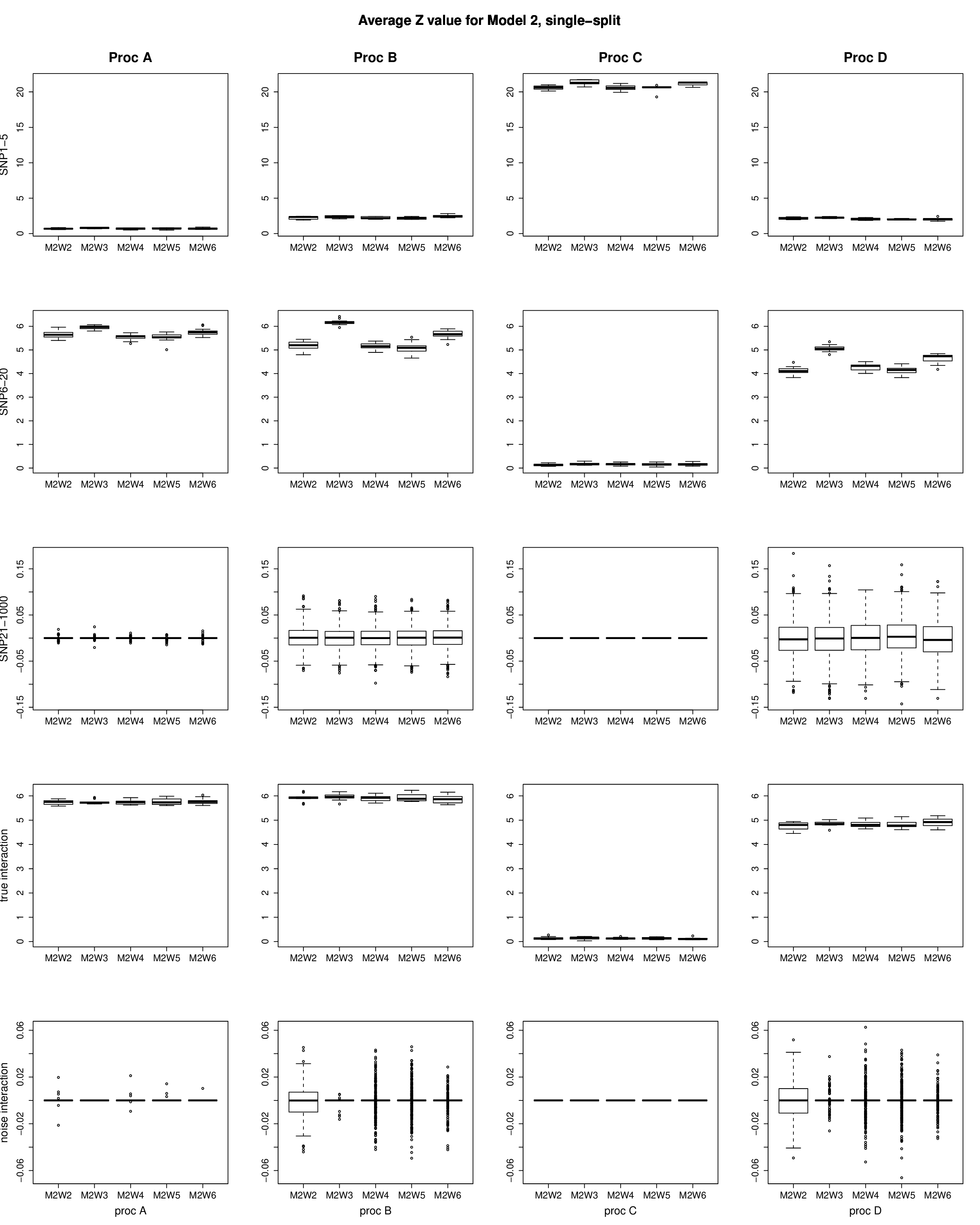}
\caption{Single-split Model 2: Average Z values for Procedure A, B, C, D with 5 W matrices. Predictors/interactions are categorized into 5 groups: active SNPs involved in interactions, active SNPs not involved in interactions, non-active SNPs, true interactions, noise interactions}\label{fig:3-3}
\end{figure*}

\begin{figure*}[!tbp]
\centering
\includegraphics[width=15cm,height=15cm]{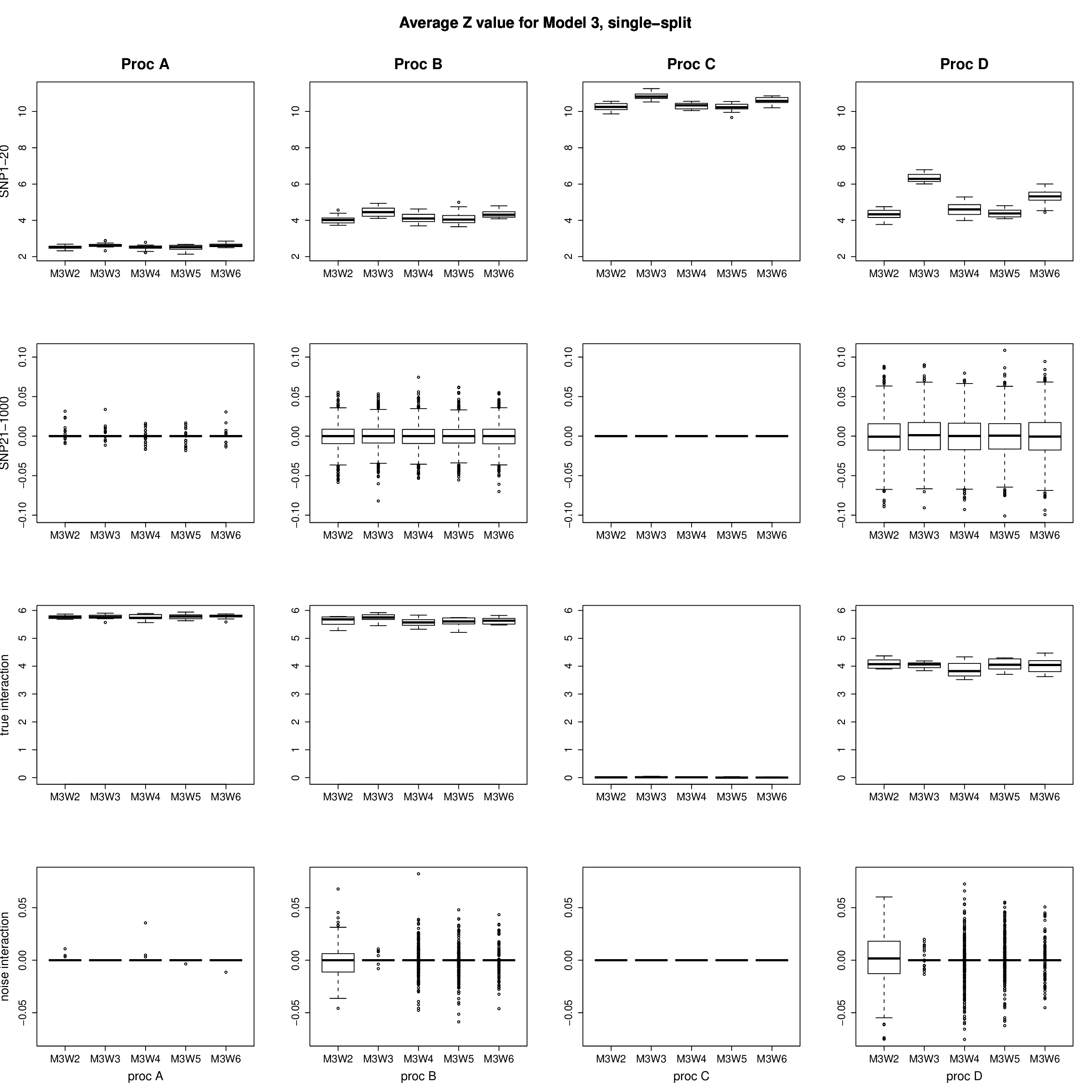}
\caption{Single-split Model 3: Average Z values for Procedure A, B, C, D with 5 W matrices. Predictors/interactions are categorized into 4 groups: active SNPs, non-active SNPs, true interactions, noise interactions}\label{fig:3-4}
\end{figure*}

We also test all the procedures using single split. The results of Z scores averaging over 100 simulations are shown in Figure \ref{fig:3-3} for Model 2 and Figure \ref{fig:3-4} for Model 3. As we can see, the observations from Figure \ref{fig:3-1} and \ref{fig:3-2} also hold in these two figures. Procedure B is the one that performs most closely to Procedure A, what should be expected if data of all cohorts were merged together and analyzed as one dataset. Also, Procedure C performs differently from Procedure A and B when selecting interactions involving main effects. This difference reflects the same pattern we observe in the multiple splits. And by adding information sharing in Procedure C (i.e. Procedure D), it performs much more closely to Procedure A, although not as close as Procedure B. 

\begin{figure*}[!tbp]
\centering
\includegraphics[width=15cm,height=18cm]{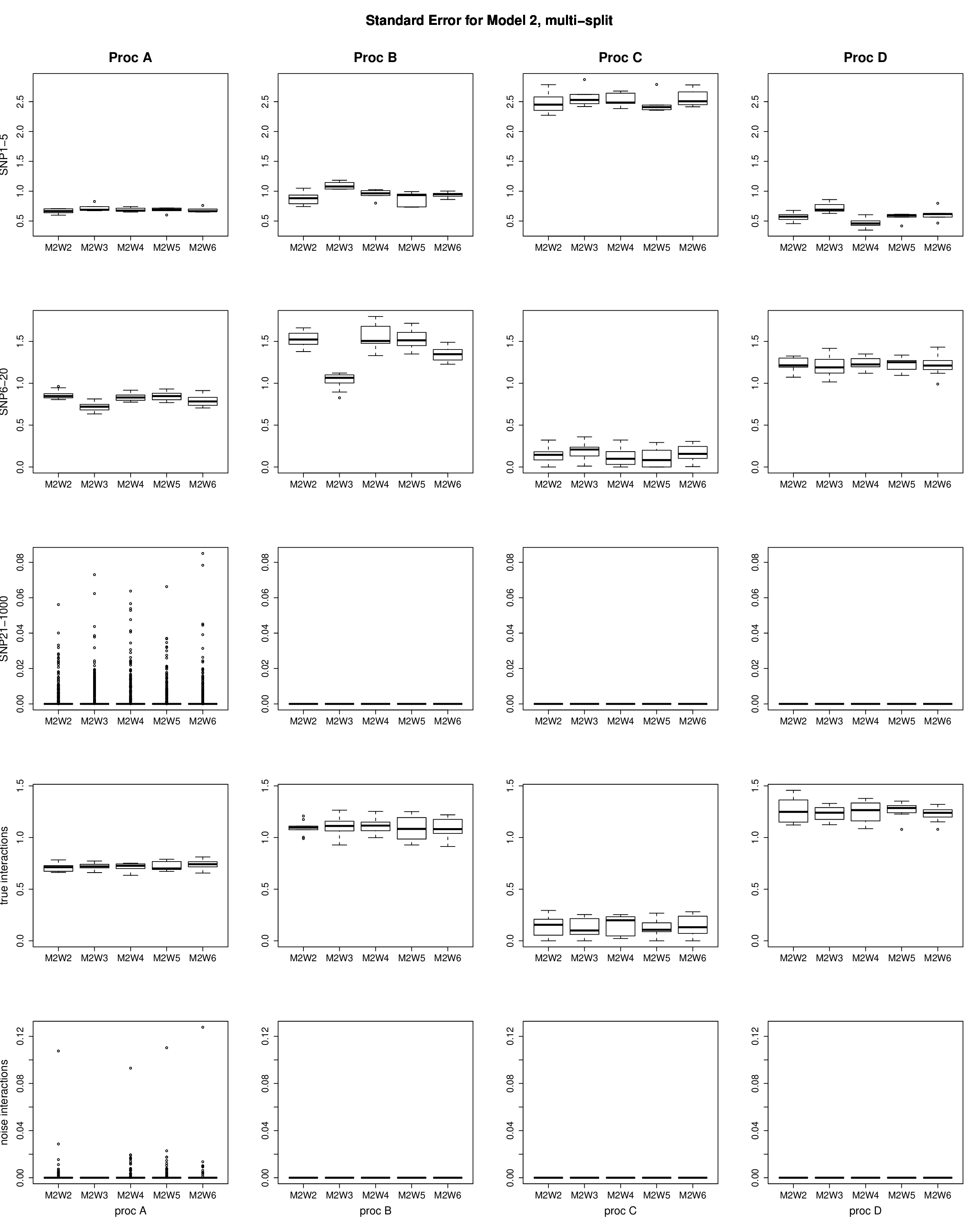}
\caption{Multi-split Model 2: Standard error of Z values for Procedure A, B, C, D with 5 W matrices. Predictors/interactions are categorized into 5 groups: active SNPs involved in interactions, active SNPs not involved in interactions, non-active SNPs, true interactions, noise interactions}\label{fig:3-5}
\end{figure*}

\begin{figure*}[!tbp]
\centering
\includegraphics[width=15cm,height=15cm]{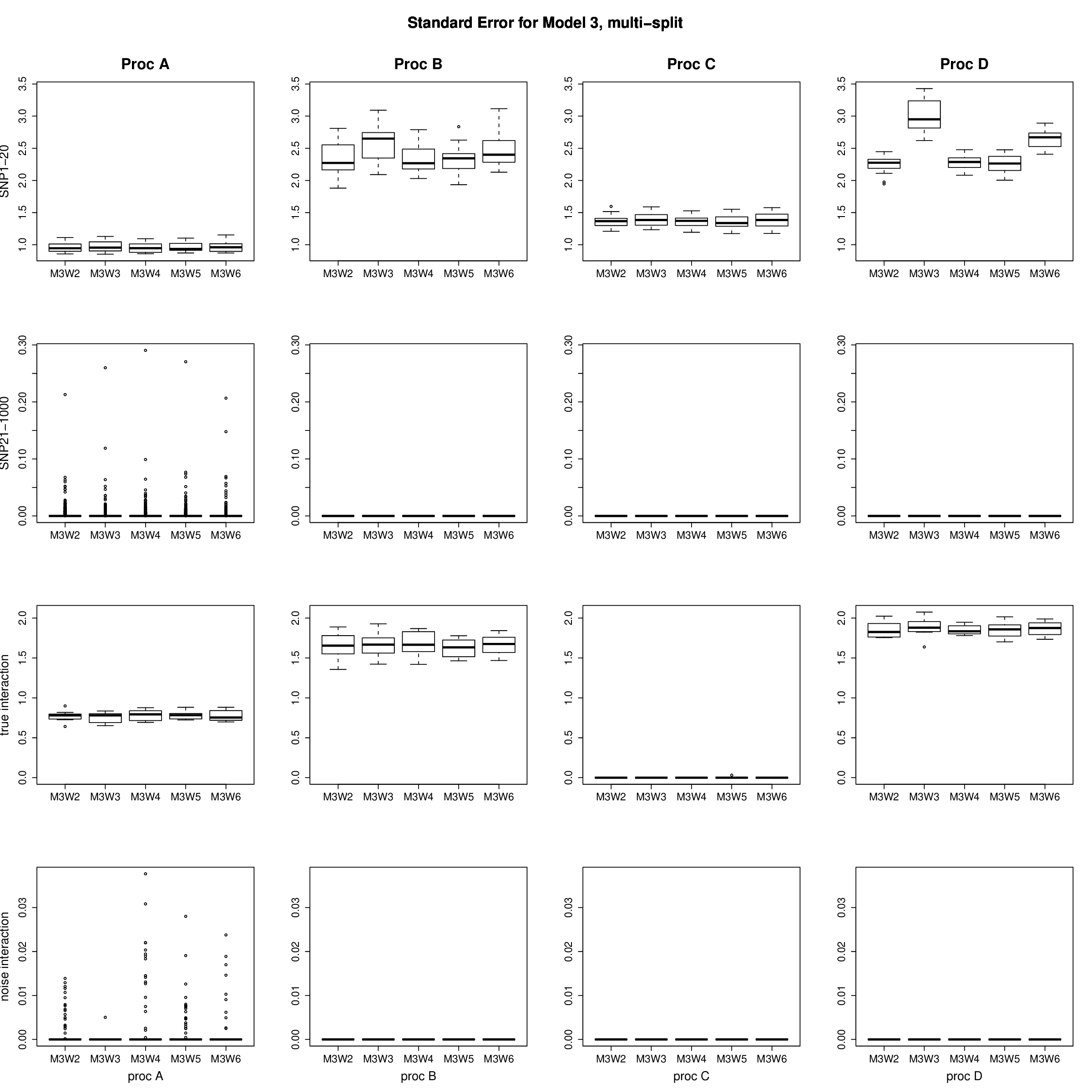}
\caption{Multi-split Model 3: Standard error of Z values for Procedure A, B, C, D with 5 W matrices. Predictors/interactions are categorized into 4 groups: active SNPs, non-active SNPs, true interactions, noise interactions}\label{fig:3-6}
\end{figure*}

\begin{figure*}[!tbp]
\centering
\includegraphics[width=15cm,height=18cm]{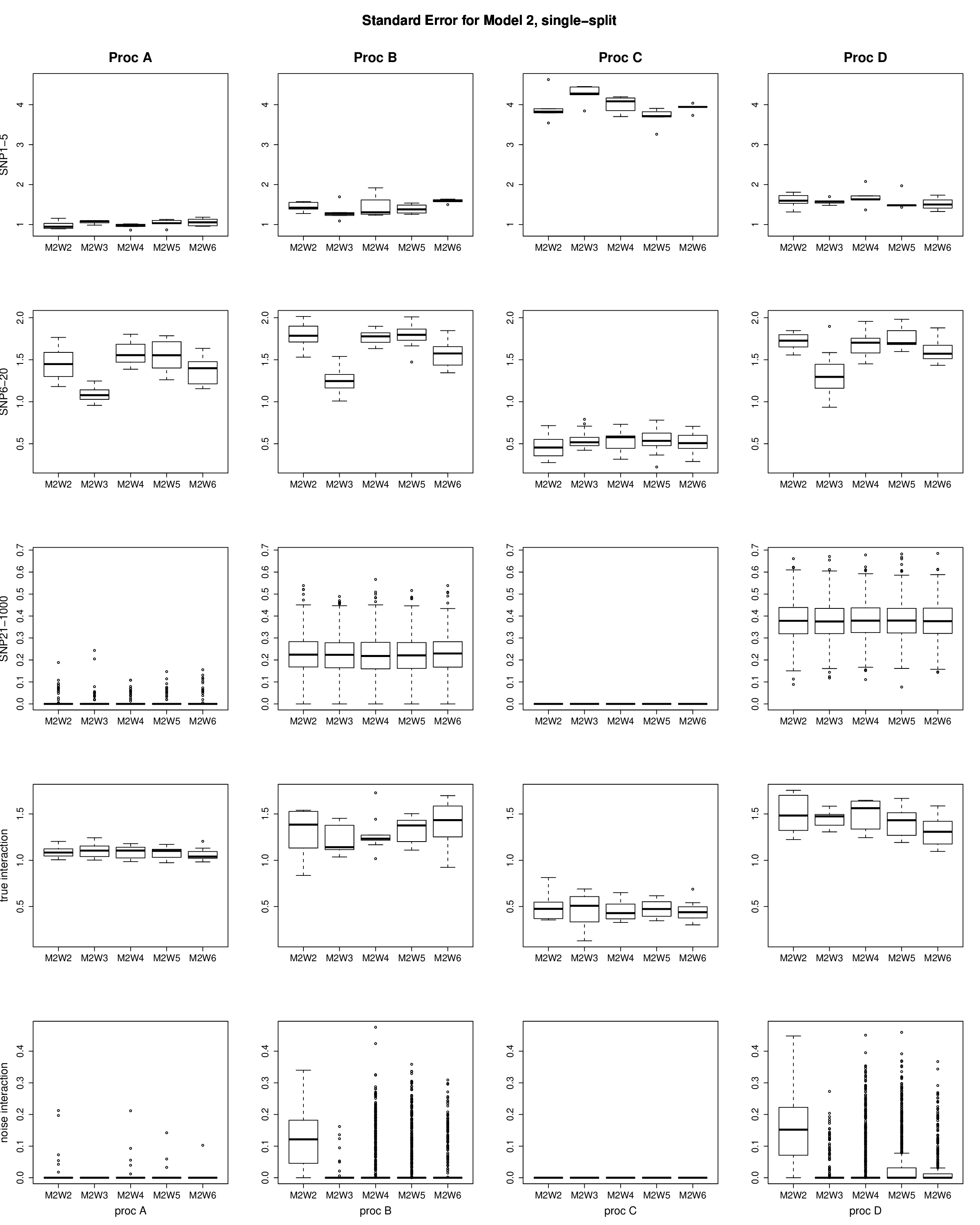}
\caption{Single-split Model 2: Standard error of Z values for Procedure A, B, C, D with 5 W matrices. Predictors/interactions are categorized into 5 groups: active SNPs involved in interactions, active SNPs not involved in interactions, non-active SNPs, true interactions, noise interactions}\label{fig:3-7}
\end{figure*}

\begin{figure*}[!tbp]
\centering
\includegraphics[width=15cm,height=15cm]{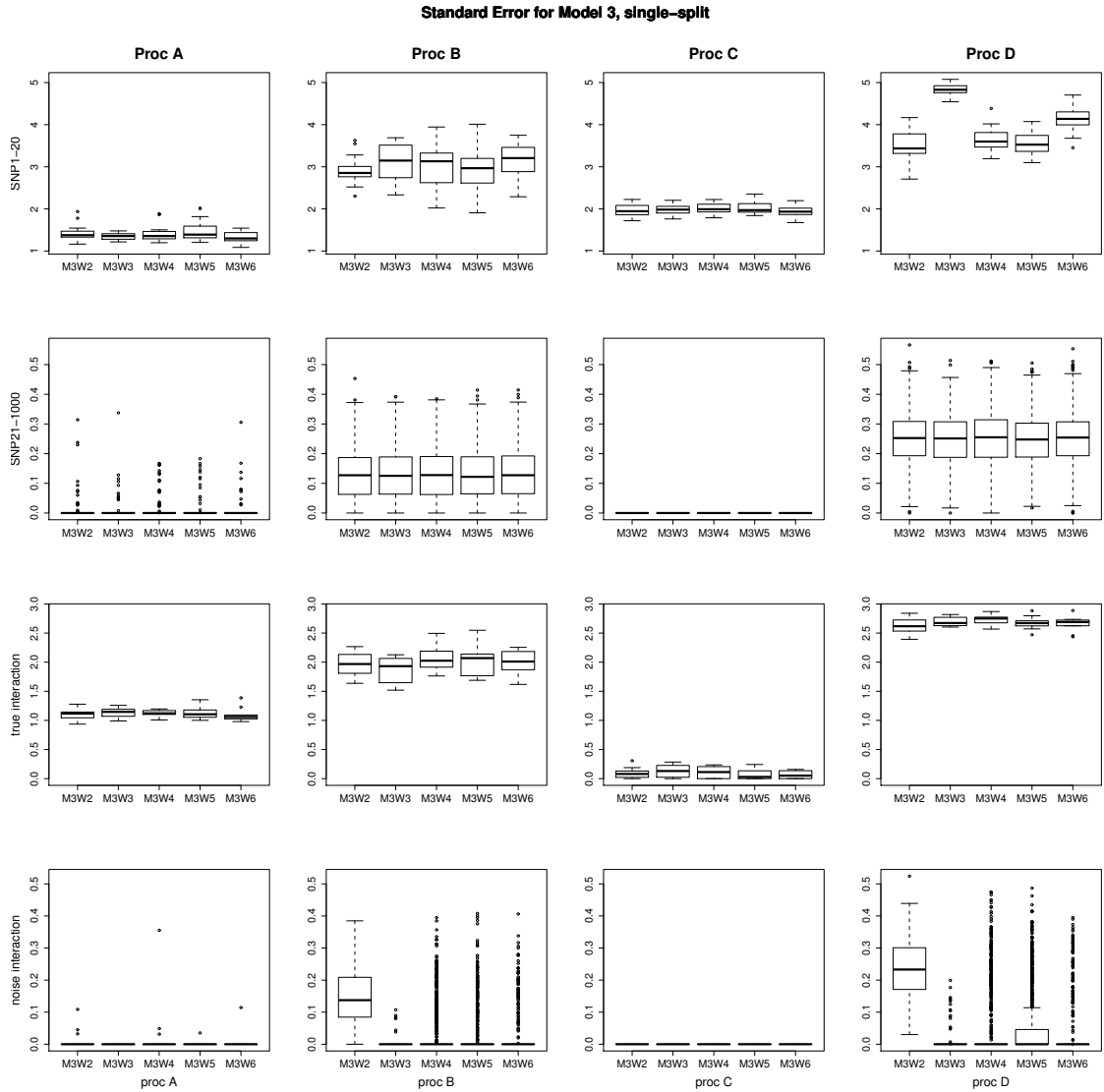}
\caption{Single-split Model 3: Standard error Z values for Procedure A, B, C, D with 5 W matrices. Predictors/interactions are categorized into 4 groups: active SNPs, non-active SNPs, true interactions, noise interactions}\label{fig:3-8}
\end{figure*}

When comparing single split vs. multiple splits, we examine the standard errors of the Z scores. As \cite{Meinshausen2009} pointed out, multiple splits method is better than single split because the result of single split depends on the arbitrary split chosen. We present the standard errors of Z scores for multiple splits in Figure \ref{fig:3-5} and \ref{fig:3-6} for Models 2 and 3, single split in Figure \ref{fig:3-7} and \ref{fig:3-8} for Models 2 and 3, respectively. By comparing Figure \ref{fig:3-5} and Figure \ref{fig:3-7}, we can see that single split has a much larger standard error of Z scores compared to multiple split, meaning that the result of single split is more variable, which is consistent with what \cite{Meinshausen2009} suggested. Comparison of Figure \ref{fig:3-6} to Figure \ref{fig:3-8} for Model 3 leads to the same conclusion.

\section{Discussion}
We examined four procedures to conduct meta-analysis using our method proposed in Chapter 2. The simulations showed that Procedure B is the one that performs most closely to the ideal case (i.e. when data are merged together and analyzed as one dataset). Modifying Procedure C to allow information sharing improves its performance to be closer to Procedure A, but still not as close as Procedure B is. 

We compared the multiple splits and single split methods in simulations. Both of them showed that Procedure B has the most similar performance as Procedure A, compared to the other two procedures. But the multiple splits method is clearly better than the single split approach, in terms of the stability of the result. 

Another interesting thing we observe here is that Procedure B is the preferred procedure in terms of performance, but is also the most practical strategy among the three procedures B, C and D. Although we already showed that multi-split is better than single split, it is a complicated process to implement multi-split in Procedure C and D. In Procedure C and D, each cohort has to split their own data $K$ times, assuming $K$ is the number of splits. The variable selection and p-value estimation are done within each cohort. This will be very complicated to coordinate among studies, and is an error-prone process. On the other hand, if we look back the algorithm of Procedure B, we can find that it is much easier to implement multi-split in Procedure B. All cohorts perform the variable selection analysis only once and report to the meta-analysis center, where the data-splitting is done $K$ times randomly. According to the $K$ assignment of groups, results of cohorts in the first group are pooled and used to create $K$ union lists of selected terms. The $K$ lists are sent to the cohorts in the second group, according to the corresponding $K$ assignments. And these cohorts are asked to perform a final model fitting. Each cohort will be asked to perform the final fitting at most $K$ times.   

We expect that the conclusions presented here should be relevant to a variety of other Lasso-based methods more generally. As we didn\rq{}t use any exclusive features of our method when developing the meta-analysis strategies, our methods are applicable generally. And it is an interesting direction for future study to apply the meta-analysis approaches to other Lasso-based method.

\cleardoublepage

\chapter{Real Data Applications}
\label{chapter:  Real Data Applications}
\thispagestyle{myheadings}

In this chapter, we apply our proposed methodology to three real data examples from the Framingham Heart Study. We will show that our method is effective in identifying potentially interesting interactions in these applications.

\section{Application to IgE Concentration}
We applied our algorithm to evaluate gene-by-gene interactions for log plasma IgE concentration, a biomarker that is often elevated in individuals with allergy to environmental allergens. An elevated plasma IgE concentration is associated with allergic diseases including asthma, allergic rhinoconjunctivitis, atopic dermatitis, and food allergy.  Although several genes influencing IgE concentrations have been identified to date, the interaction among these genes or others yet to be identified to be important players have not been studied \cite{Granada2012}.

We sought to investigate gene-by-gene effects on log IgE concentration in the Framingham Heart Study (FHS) cohorts.  Participants from the town of Framingham, Massachusetts have been recruited in these studies starting in 1948, and have been followed over the years for the development of heart disease and related traits, including pulmonary function and allergic response measured by IgE concentration.  Our analyses include 6975 participants, 441 from the original cohort recruited in 1948, an additional 2848 from the Offspring cohort recruited in 1971, and finally 3686 participants from the third generation cohort initiated in 2002.  A recent genome-wide association study on Framingham participants identified new genetic loci associated with plasma total IgE concentrations \cite{Granada2012}. We are interested in looking at GxG interactions associated with IgE concentration, as an illustration of our methodology.

\subsection{Preliminaries}
Genotypes were from Affymetrix 500K and  MIPS 50K arrays, with imputation performed using HapMap 2 European reference panel \cite{li2006mach}.  Dosage genotypes (expected number of minor alleles) were used in our analysis, although the software implementation of the  ~\cite{wu2009genome} approach (Mendel) required genotypes to be coded as $0$, $1$ or $2$ and could not handle dosage.  Therefore, in our analysis using Mendel, for each individual we used the genotype with the highest posterior probability at each SNP. We analyze the natural logarithm of plasma total IgE concentrations as our phenotype (i.e., $Y$) adjusted for smoking status (current, former and amount of life time smoking in terms of pack-years), age, sex, and cohort of origin. A total of 6975 participants (3209 men and 3766 women) age 19 and older had good quality genotypes and were included in our analysis. Familial relationship was ignored when applying our algorithm and the \cite{wu2009genome} approach, but we subsequently applied linear mixed effect models to account for familial correlation to obtain estimates of effect sizes.

Some pre-processing was used to select a set of SNPs to include in our analysis.  First, we attempted to map each of the 2,411,590 genotyped and imputed SNPs in the dataset to a reference gene containing it.  If no such gene was available, then we mapped the SNP to the closest reference gene within 60 kilobases of the SNP, if available. Otherwise, the SNP was excluded.  After establishing this mapping between genes and SNPs, some genes were found to include multiple SNPs. We kept only one SNP for each gene, selecting in each case the SNP most significantly associated with the phenotype, based on a  linear mixed effect regression. As a result, the SNPs in the final data set have low linkage disequilibrium (correlation) and a unique SNP-to-gene correspondence. (As we will show in Section~4.1.4, our results reported below are fairly robust to modest amounts of disequilibrium in these data.)

The final data set has $17,025$ SNPs/Genes. We used the KEGG (Kyoto Encyclopedia of Genes and Genomes) pathway database to build our $W$ matrix, following the steps described in the Methods section of Chapter 2. The KEGG pathway database has a total of $72,354$ genes and $5,268$ unique genes, resulting in $479,066$ interactions allowed in our $W$ matrix.

\subsection{Results Using KEGG to Construct W Matrix}
For our analysis on $17,025$ SNPs, we chose to look for 10 main effects, although we allowed the algorithm to terminate after selecting 10 plus or minus one main effect, resulting in 9 main effects selected in the current analysis.  The parameter $c$ was set to $0.1$ which, based on an average estimated SNP variance of 0.27 for these data, corresponds to $r\approx 2.7$.  Six interactions were found in our approach, yielding a model with a total of $15=9+6$ variables.  In order to calibrate our results with those from the stage-wise procedure of~\cite{wu2009genome}, as implemented in Mendel, the latter was run to select $9$ variables in the first stage (i.e., fitting only main effects), and then $15$ variables in the second stage (i.e., fitting both main effects and interactions, selected from among the $9$ SNPs resulting from the first stage).  This process produced a final model with 9 main effects and 6 interactions.  In terms of computing time, our analysis ran in roughly 5 minutes on our cluster Linga, equipped with 2 Intel Xeon CPUs E5345 @ 2.33GHz with 4 cores each and 16 GB / 32 GB of RAM for each node (the job was submitted to one node and used one core), while the analysis in Mendel ran in roughly 2 minutes.  Given that our method evaluates $479,066/55\approx 8710$ times more potential interactions than Mendel, the observed trade-off between computing time and number of possible interactions being evaluated appears to be quite reasonable.

The results from our proposed method and from the stage-wise procedure are shown in the left and right, respectively, of Table \ref{tab:4-1}.  The estimates of effect size and the ranks are from the linear mixed effect model for the final model after variable selection procedure, for both methods.  Genes previously found in a GWAS of these FHS data \cite{Granada2012} are indicated with an asterisk in the table.  In our approach, four of the six interaction pairs involved human leukocyte antigen (HLA) genes, which encode antigen-presenting cell-surface proteins that are key regulators of the immune response. The other two interactions identified were among genes both previously associated with log IgE concentrations \cite{Granada2012}. In contrast, Mendel did not detect any interactions among genes in the HLA regions or among pairs of previously associated genes.

\begin{table}[h]
\caption{Results of application to IgE concentration data.}
\footnotesize{
\centering
\begin{tabular}{cccc|cccc}
\multicolumn{4}{c}{Network-Guided Sparse Regression} & \multicolumn{4}{c}{Mendel analysis}\\
\hline
Gene1 & Gene2 & t-value & Found & Gene1 & Gene2 & t-value & Found \\
\hline
FCER1A	&		&	-5.6441	&	*	&	LRP1	&		&	4.7084	&		\\
MPP6	&		&	4.4184	&		&	SNF1LK2	&		&	4.3969	&		\\
STAT6	&		&	-4.2453	&	*	&	EMID2	&		&	-4.1795	&		\\
IL13	&		&	4.0073	&	*	&	RAB3C	&		&	3.8585	&		\\
LRP1	&		&	3.7072	&		&	HLA-DQA2	&		&	3.6883	&	*	\\
HLA-DPB1	&	HLA-DQA2	&	1.6314	&		&	FCER1A	&		&	-2.8098	&	*	\\
FCER1A	&	HLA-DQA2	&	1.4193	&		&	HLA-DPB1	&		&	2.1346	&		\\
HLA-G	&		&	1.3657	&	*	&	LOC441108	&		&	1.9687	&		\\
HLA-DPB1	&		&	1.1655	&		&	LOC441108	&	DDX1	&	1.7449	&		\\
HLA-A	&		&	0.8442	&	*	&	LRP1	&	DDX1	&	-1.6417	&		\\
FCER1A	&	IL13	&	0.6318	&		&	FCER1A	&	SNF1LK2	&	-1.5967	&		\\
HLA-DQA2	&		&	0.4590	&	*	&	DDX1	&	SNF1LK2	&	-1.4047	&		\\
HLA-A	&	HLA-DPB1	&	0.4318	&		&	DDX1	&		&	-1.1802	&		\\
HLA-G	&	HLA-A	&	-0.2813	&		&	HLA-DPB1	&	EMID2	&	0.8505	&		\\
HLA-G	&	HLA-DQA2	&	0.0678	&		&	HLA-DPB1	&	LOC441108	&	-0.8076	&		\\
\hline
\end{tabular}{ Terms are ranked based on absolute t value, * in found column represents the genes that were found in publication.}
}
\label{tab:4-1}
\end{table}

From a biological perspective, a number of the interactions discovered by our method are of nontrivial potential interest. The MHC class I antigens HLA-A, -B, and -C are involved with cell-mediated immunity targeting cells expressing proteins produced intracellularly, for example by viruses, while the MHC class II antigens HLA-DP, -DQ, and -DR play key roles with humoral immunity, including the production of IgE antibodies directed against environmental allergens (\cite{Klein2000}). HLA-G is a nonclassical MHC class I antigen that may have immunomodulatory effects through actions on natural killer cells, T lymphocytes, and antigen-presenting cells (\cite{Carosella2008}). Genetic variants in these different classes of HLA genes --- each class influencing a different but interconnected aspect of immune function --- could well interact to influence the risk of developing IgE dysregulation and allergy. The observed interaction between SNPs in the FCER1A and IL13 genes may reflect a number of mechanisms. For example, a genetic variant causing increased expression of Fc$\varepsilon$RI$\alpha$ on mast cells would lead to increased antigen-induced activation of these cells, which would consequently produce more IL-13 (\cite{burd1995activated}), leading to more class switch recombination and IgE production. Genetic variation of Fc$\varepsilon$RI$\alpha$ on classical antigen-presenting cells may also promote Th2 cell activation (\cite{potaczek2009genetic}) with consequent IL-13 release. Thus, SNPs in these two genes in the same pathway leading to increased IgE production could have synergistic effects.  Overall, identification of these interactions may help identify the children at highest risk for developing allergy, possibly helping focus interventions to prevent allergy, and may provide new insights into the genetic basis and mechanisms of allergy.

\subsection{Using Other Biological Databases}
In the previous analysis, we used KEGG pathway database to construct our W matrix to incorporate in the penalty. In order to see if there are additional interactions, we use two other biological databases to construct the W matrix, GO (Gene Ontology) and HPRD (Human Protein Reference Database) protein-protein interaction database.

KEGG database provides information on pathway and groups genes according to the biochemical pathways they are involved in. GO database groups genes according to biological functions (we used biological process collection). To construct the W matrix from GO, we use gene sets with less than 300 genes so that we include gene sets with more specific biological functions, and follow the steps described in Chapter 2 Section 2.1.

On the other hand, HPRD database provides the information on protein-protein interactions and the data consists of pairs of interacting genes. We use the gene pairs as indicators of possible interactions and construct the $W$ matrix by assigning the indicators to the elements of the matrix. 

For these two analyses, we used a more liberal set of parameters. We set to choose 30 main effects and the parameter $c$ was set to $0.01$ so that we could choose a fair amount of interactions. Here the main purpose is to see if there are any potential interesting interactions in addition to the main findings we obtained using the KEGG pathway database. 

\begin{table}[h]
\caption{Results of application to IgE concentration using GO to construct W matrix.}
\centering
\begin{tabular}{cccc}
\multicolumn{4}{c}{Using GO to construct W matrix } \\
\hline
Gene1 & Gene2 & t-value & Main effects Found  \\
\hline
FCER1A	&		&	-7.5568	&	*	\\
SNFT	&		&	-5.2928	&		\\
HLA-DPB1	&		&	5.1444	&		\\
DDX1	&		&	-4.6712	&		\\
CDH11	&		&	4.4932	&		\\
SNF1LK2	&		&	4.4436	&		\\
EMID2	&		&	-4.3246	&		\\
IL4R	&		&	-4.2640	&		\\
MPP6	&		&	4.2370	&		\\
ANKS4B	&		&	4.2282	&		\\
LOC441108	&		&	4.1518	&		\\
PPP2R2B	&		&	4.1266	&		\\
CST5	&		&	-4.0151	&		\\
C6orf85	&		&	3.9109	&		\\
TLN2	&		&	3.8613	&		\\
HLA-DRB1	&		&	-3.8140	&		\\
LRP1	&		&	3.7417	&		\\
STAT6	&		&	-3.7026	&	*	\\
HAND1	&		&	-3.6482	&		\\
GREM2	&		&	-3.5610	&		\\
AMDHD1	&		&	-3.4063	&		\\
RAB3C	&		&	3.3114	&		\\
CMA1	&		&	-3.1356	&		\\
TRPM6	&		&	3.0290	&		\\
IL13	&		&	2.4194	&	*	\\
HLA-G	&	HCP5	&	1.9078	&		\\
HLA-A	&		&	1.5245	&	*	\\
IL13	&	HCP5	&	1.4447	&		\\
HCP5	&		&	1.1852	&		\\
IL13	&	HLA-G	&	1.1163	&		\\
HLA-DQA2	&		&	0.9184	&	*	\\
HLA-G	&		&	-0.0546	&	*	\\
\hline
\end{tabular}
\label{tab:4-2}
\end{table} 

The result of the analysis using GO database is shown in Table \ref{tab:4-2}. Using a different W matrix, we identified the same main effects found in the published GWAS results of these FHS data \cite{Granada2012}, indicated with an asterisk in the table. One of the interactions found is of potential interest given than both genes (IL13 and HLA-G) have previously been identified.

\begin{table}[h]
\caption{Results of application to IgE concentration using HPRD PPI to construct W matrix.}
\centering
\begin{tabular}{cccc}
\multicolumn{4}{c}{Using HPRD PPI to construct W matrix } \\
\hline
Gene1 & Gene2 & t-value & Main effects Found  \\
\hline
FCER1A	&		&	-7.535390555	&	*	\\
SNFT	&		&	-5.27318176	&		\\
HLA-DPB1	&		&	5.137494999	&		\\
IL13	&		&	4.803813165	&	*	\\
DDX1	&		&	-4.703033381	&		\\
CDH11	&		&	4.536470905	&		\\
SNF1LK2	&		&	4.476663127	&		\\
EMID2	&		&	-4.340102709	&		\\
MPP6	&		&	4.24129508	&		\\
ANKS4B	&		&	4.177733918	&		\\
LOC441108	&		&	4.118810407	&		\\
STAT6	&		&	-4.070155021	&	*	\\
PPP2R2B	&		&	4.068292492	&		\\
CST5	&		&	-4.04076742	&		\\
C6orf85	&		&	3.901504818	&		\\
TLN2	&		&	3.787433938	&		\\
LRP1	&		&	3.7574075	&		\\
HAND1	&		&	-3.663633263	&		\\
GREM2	&		&	-3.546774073	&		\\
AMDHD1	&		&	-3.42627879	&		\\
IL4R	&		&	-3.425851751	&		\\
RAB3C	&		&	3.414224153	&		\\
HCP5	&		&	3.369049934	&		\\
CMA1	&		&	-3.111652119	&		\\
TRPM6	&		&	3.012148241	&		\\
STAT6	&	IL4R	&	1.668298955	&		\\
HLA-A	&		&	1.504116975	&	*	\\
HLA-G	&		&	1.491234609	&	*	\\
HLA-DRB1	&		&	-1.174091022	&		\\
IL13	&	IL4R	&	-1.148257656	&		\\
HLA-DRA	&		&	-0.946271115	&		\\
HLA-DQA2	&		&	0.899191618	&	*	\\
HLA-DRA	&	HLA-DRB1	&	-0.258048129	&		\\
\hline
\end{tabular}
\label{tab:4-3}
\end{table} 

The result of the analysis using HPRD PPI (protein-protein interactions) database is shown in Table \ref{tab:4-3}. Here we again found the same main effects reported in the published GWAS result of the FHS data \cite{Granada2012}, indicated with an asterisk in the table. 
Among the interactions found in this set of analysis, there is one in the HLA region, and others involving the Interleukin genes IL13 and IL4R.

\subsection{Effect of Linkage Disequilibrium}
Another important issue to investigate is the effect of linkage disequilibrium (LD) among SNPs, because regular Lasso methods assume unrelated predictors and do not account for correlations among predictors. In our analysis of log plasma IgE concentrations, we select 1 SNP per gene to decrease correlation between SNPs. Here we perform three more analyses by selecting 1 SNP, 3 SNPs or 5 SNPs per gene, respectively. $W$ matrices for the analyses are constructed using the KEGG pathway database. All three analyses are restricted to Chromosome 6, i.e., where the previously identified main effects and interactions are concentrated, since our focus here is not so much on the discovery of additional interactions, but rather on assessing the robustness of our previous findings when some degree of LD exists. We set the tuning parameters the same way as we did when using the KEGG pathway database in Section~4.1.2 for all three analyses.  Specifically, $\lambda_1$ is set to select 10 main effects and $c$ is set to 0.1.

The results from these analyses are shown in Table \ref{tab:realdata}.  The two analyses with 3 SNPs per gene and with 5 SNPs per gene have selected exactly the same main effects and interactions in terms of genes, so we present their results as one, in comparison to the analysis with 1 SNP per gene. From the table we can see that the one-SNP/gene analysis and the 3\&5-SNP/gene analyses selected many of the same main effects/interactions at the gene level.  Furthermore, there is substantial instances of gene-level main effects and interactions found in the 3\& 5-SNP analyses, due to the selection of multiple SNPs per gene. More specifically, in terms of main effects, the one-SNP/gene analysis found 11 main effects, while the 3\& 5-SNP/gene analyses found 7 unique main effects (11 non-unique), 6 of which were among those found by the one-SNP/gene analysis.  Similarly, the one-SNP/gene analysis found 14 interactions, and the 3\& 5-SNP/gene, 15 interactions (32 non-unique), with 6 interactions in common.  Combining main effects and interactions, the corresponding Jaccard coefficient was $12/23=0.522$. (The Jaccard coefficient measures similarity between two sample sets, and is defined as the size of the intersection divided by the size of the union of the sample sets.)

In summary, the effect of modest LD among SNPs has not seemed to substantially affect the selection of terms in this example.

\definecolor{purple}{rgb}{0.7,0,1}
\begin{table}[h]
\centering
{\footnotesize
\begin{tabular}{ccc|ccc}
\multicolumn{3}{c}{Analysis with 1 SNP per gene} & \multicolumn{3}{c}{Analysis with 3 SNP, or 5 SNP per gene} \\
\hline
Gene1	&	Gene2	&	t-value	&	Gene1	&	Gene2	&	t-value \\
\hline
\color{blue}GABRR2	&		&	-3.8654	&	\color{red}C6orf85	&		&	3.8465	\\
\color{red}C6orf85	&		&	3.661	&	\color{purple}HLA-DPB1	&	\color{purple}HLA-DPB1	&	-2.5109	\\
\color{blue}HCP5	&		&	2.7668	&	\color{red}HLA-A	&	\color{red}HLA-DPB1	&	-2.5054	\\
\color{blue}ITPR3	&		&	-2.7008	&	\color{red}HLA-DPB1	&		&	2.2225	\\
\color{blue}HLA-G	&	\color{blue}HLA-DRB1	&	-2.1174	&	\color{purple}HLA-DPB1	&	\color{purple}HLA-DPB1	&	2.221	\\
\color{red}HLA-G	&		&	1.7588	&	\color{purple}HLA-G	&	\color{purple}HLA-DQB1	&	-2.1499	\\
\color{blue}HLA-DRA	&	\color{blue}HLA-B	&	-1.6056	&	\color{red}HLA-B	&		&	1.9161	\\
\color{red}HLA-DQA2	&	\color{red}HLA-DPB1	&	1.3734	&	\color{red}HLA-G	&	\color{red}HLA-DQA2	&	-1.8227	\\
\color{red}HLA-DPB1	&		&	1.332	&	\color{red}HLA-G	&	\color{red}HLA-DQA2	&	1.5974	\\
\color{blue}HLA-DRA	&	\color{blue}HLA-A	&	1.1896	&	\color{purple}HLA-A	&	\color{purple}HLA-DQA2	&	1.5698	\\
\color{red}HLA-G	&	\color{red}HLA-B	&	-0.8313	&	\color{purple}HLA-A	&	\color{purple}HLA-DQA2	&	1.5636	\\
\color{blue}HLA-DRA	&		&	-0.755	&	\color{purple}HLA-DPB1	&	\color{purple}HLA-DPB1	&	1.3393	\\
\color{blue}HLA-DQA2	&	\color{blue}HLA-DRB1	&	-0.7169	&	\color{purple}HLA-DQA2	&	\color{purple}HLA-DQB1	&	-1.3266	\\
\color{red}HLA-B	&	\color{red}HLA-DPB1	&	-0.6759	&	\color{red}HLA-DQA2	&	\color{red}HLA-DPB1	&	-1.1567	\\
\color{blue}HLA-A	&	\color{blue}HLA-DRB1	&	0.6391	&	\color{red}HLA-DQA2	&	\color{red}HLA-DPB1	&	1.053	\\
\color{blue}HLA-DRA	&	\color{blue}HLA-DQA2	&	0.6107	&	\color{red}HLA-A	&	\color{red}HLA-G	&	1.0475	\\
\color{red}HLA-G	&	\color{red}HLA-DQA2	&	-0.6104	&	\color{red}HLA-A	&	\color{red}HLA-G	&	1.0326	\\
\color{blue}HLA-DRA	&	\color{blue}HLA-DRB1	&	-0.2642	&	\color{red}HLA-DPB1	&		&	-0.8669	\\
\color{red}HLA-DPB1	&	\color{red}HLA-A	&	0.2427	&	\color{red}HLA-A	&	\color{red}HLA-DPB1	&	-0.8565	\\
\color{red}HLA-A	&		&	0.0869	&	\color{purple}HLA-G	&	\color{purple}HLA-DPB1	&	-0.8522	\\
\color{blue}HLA-B	&	\color{blue}HLA-DQA2	&	0.0459	&	\color{purple}HLA-A	&	\color{purple}HLA-A	&	-0.8375	\\
\color{blue}HLA-DRB1	&		&	-0.0416	&	\color{red}HLA-DQA2	&		&	-0.7843	\\
\color{red}HLA-G	&	\color{red}HLA-A	&	-0.033	&	\color{purple}HLA-A	&	\color{purple}HLA-DQB1	&	0.7532	\\
\color{red}HLA-B	&		&	-0.0313	&	\color{red}HLA-B	&	\color{red}HLA-DPB1	&	0.7229	\\
\color{red}HLA-DQA2	&		&	0.0261	&	\color{red}HLA-G	&	\color{red}HLA-B	&	0.6769	\\
	&		&		&	\color{purple}HLA-A	&	\color{purple}HLA-DQB1	&	-0.6027	\\
	&		&		&	\color{red}HLA-A	&	\color{red}HLA-DPB1	&	-0.4971	\\
	&		&		&	\color{red}HLA-A	&	\color{red}HLA-G	&	-0.4673	\\
	&		&		&	\color{red}HLA-G	&	\color{red}HLA-B	&	-0.4622	\\
	&		&		&	\color{red}HLA-A	&		&	-0.4584	\\
	&		&		&	\color{purple}HLA-G	&	\color{purple}HLA-G	&	0.4567	\\
	&		&		&	\color{purple}HLA-DQB1	&		&	-0.3608	\\
	&		&		&	\color{red}HLA-B	&	\color{red}HLA-DPB1	&	0.3006	\\
	&		&		&	\color{red}HLA-DPB1	&		&	0.2888	\\
	&		&		&	\color{red}HLA-A	&		&	0.2234	\\
	&		&		&	\color{purple}HLA-G	&	\color{purple}HLA-DQB1	&	0.2223	\\
	&		&		&	\color{red}HLA-A	&	\color{red}HLA-DPB1	&	-0.1681	\\
	&		&		&	\color{red}HLA-A	&	\color{red}HLA-DPB1	&	-0.1242	\\
	&		&		&	\color{red}HLA-A	&	\color{red}HLA-DPB1	&	-0.1175	\\
	&		&		&	\color{red}HLA-G	&		&	-0.0746	\\
	&		&		&	\color{red}HLA-A	&	\color{red}HLA-G	&	-0.0638	\\
	&		&		&	\color{purple}HLA-B	&	\color{purple}HLA-DQB1	&	-0.0413	\\
	&		&		&	\color{red}HLA-G	&		&	-0.0276	\\
\hline
\end{tabular}
}
\caption{Results of application to IgE concentration data, with varying LD among SNPs. Terms are ranked based on absolute t value. \color{blue}Blue: genes / interactions found only in one-SNP/gene analysis; \color{purple}Purple: genes / interactions found only in 3\&5-SNPs/gene analyses; \color{red}Red: genes / interactions found in all three analyses.} \label{tab:realdata}
\end{table}

\section{Application to CRP Serum Levels}
\subsection{Background}
We apply our method to evaluate gene-by-gene interactions influencing C-reactive protein serum levels. C-reactive protein (CRP) is a general marker of systemic inflammation. High CRP levels are associated increased risks of mortality and major disease including diabetes mellitus, hypertension, coronary heart disease and stroke. CRP is also a heritable marker of chronic inflammation that is strongly associated with cardiovascular disease. Eighteen loci associated with CRP levels have been identified to be associated with CRP levels \cite{Dehghan2011}. But interactions have not been identified yet and we are interested in looking at gene-by-gene interaction associated with CRP levels using the Framingham Heart Study data.

This analysis has 6899 participants which include 3221 men and 3678 women with age 19 and older, 3852 from the Offspring cohort and 3047 from the third generation cohort. We analyze the natural logarithm of CRP serum levels as our phenotype, adjusted for age and sex.

We ignore familial relationship when applying our algorithm and account for the relatedness subsequently using a linear mixed effect model. We perform the same pre-processing to select a set of SNPs such that only one SNP is kept for each gene and the SNPs in the final data set have low linkage disequilibrium and a unique SNP-to-gene correspondence. The final data set has 17,569 SNPs/Genes. As we already saw from the first application, KEGG was more effective in finding interactions than GO and HPRD databases.  So for this application, we used only KEGG pathway database to construct W matrix, which allowed 499,687 interactions to be evaluated by the algorithm.

\subsection{Results}
For this analysis, we choose to look for 30 main effects and set parameter $c$ to $0.01$. We allow the algorithm to terminate after selecting 30 plus or minus one main effects. The analysis selects 29 main effects and 4 interactions. The selected terms are ranked according to their absolute t values in the final fitting, and shown in Table \ref{tab:4-4}. Genes previously identified in \cite{Dehghan2011} are indicated with an asterisk in the table. We find 6 main effects that were previously identified \cite{Dehghan2011}, one (out of 4) interaction in genes both previously associated with log CRP levels, and another two interactions involving one of the previously reported genes.

\begin{table}[h]
\caption{Results of application to CRP.}
\centering
\begin{tabular}{cccc}
\multicolumn{4}{c}{Network Guided Sparse Regression for CRP } \\
\hline
Gene1 & Gene2 & t-value & Main effects Found  \\
\hline
CRP	&		&	-8.018683534	&	*	\\
LEPR	&		&	-5.00387017	&	*	\\
FLJ43860	&		&	-4.363210157	&		\\
RAD23B	&		&	4.250179466	&		\\
C9orf30	&		&	-4.142057122	&		\\
ETAA1	&		&	4.087399338	&		\\
RGS6	&		&	4.051728602	&	*	\\
IFLTD1	&		&	-4.027429023	&		\\
TDRD10	&		&	-3.955953323	&		\\
TMEM132D	&		&	3.95507514	&		\\
MCART2	&		&	3.948868622	&		\\
SCN3B	&		&	-3.926474997	&		\\
SLC24A4	&		&	-3.921758746	&		\\
GCKR	&		&	3.891738579	&	*	\\
VIT	&		&	3.881651071	&		\\
GDNF	&		&	-3.737491061	&		\\
NR3C2	&		&	3.708207987	&		\\
NEBL	&		&	3.648881857	&		\\
NOTCH4	&		&	-3.5771008	&		\\
PDE8B	&		&	3.557864712	&		\\
COPS5	&		&	3.503856221	&		\\
MANSC1	&		&	3.238662051	&		\\
NHLRC1	&		&	2.686591709	&		\\
IL1R2	&		&	-2.552375629	&		\\
UGT3A2	&		&	-2.521693083	&		\\
HNF1A	&		&	-2.493997014	&	*	\\
SKP2	&		&	1.799122844	&		\\
IL6R	&		&	-1.71177651	&	*	\\
OASL	&		&	-1.657445141	&		\\
LEPR	&	IL6R	&	1.262566345	&		\\
IL6R	&	IL1R2	&	-0.861904991	&		\\
LEPR	&	IL1R2	&	0.854238003	&		\\
NHLRC1	&	SKP2	&	-0.079797391	&		\\
\hline
\end{tabular}
\label{tab:4-4}
\end{table} 

\section{Application to Fasting Glucose}
\subsection{Background}
Fasting Glucose is commonly measured to detect type II diabetes and is one of the criteria that defines type II diabetes. Some new genetic loci have been identified to be associated with fasting glucose in a recent meta-analysis \cite{dupuis2010new,Manning2012}. As a third application of our method, we are interested in investigating interactions among these newly discovered genes using Framingham Heart Study.

This analysis includes 6479 participants (2981 men and 3498 women, with age 19 and older), 2766 from the Offspring cohort and 3713 from the third generation cohort. We analyze the fasting glucose as our phenotype adjusted for age, sex, bmi and cohort of origin. 

We again ignore familial relationship when applying our algorithm and account for relatedness in subsequent analysis using linear mixed effect model. We perform the same pre-processing to select a set of SNPs such that only one SNP was kept for each gene and the SNPs in the final data set have low linkage disequilibrium and a unique SNP-to-gene correspondence. The final data set has 17,026 SNPs/Genes. For this application, we also used only KEGG pathway database to construct W matrix, which allowed 479,252 interactions to be evaluated by the algorithm.

\subsection{Results}
For the analysis of fasting glucose, we chose to look for 30 main effects and set parameter $c$ to $0.01$. We allowed the algorithm to terminate after selecting 30 plus or minus one main effects. The analysis selected 31 main effects and 6 interactions. Results of selected terms are ranked according to their absolute t values in the final fitting, as shown in Table \ref{tab:4-5}. Gene previously found in \cite{dupuis2010new} are indicated with an asterisk in the table. We found three main effects that were previously identified, one (out of six) interaction in genes both previously associated with fasting glucose \cite{dupuis2010new} and another interaction involving one of the previously reported genes.

\begin{table}[h]
\caption{Results of application to Fasting Glucose.}
\centering
\begin{tabular}{cccc}
\multicolumn{4}{c}{Network Guided Sparse Regression for Fasting Glucose } \\
\hline
Gene1 & Gene2 & t-value & Main effects Found  \\
\hline
MTNR1B	&		&	6.597569199	&	*	\\
CDKL1	&		&	4.910601399	&		\\
TTYH2	&		&	4.863204072	&		\\
C1orf201	&		&	4.637364714	&		\\
ABCB11	&		&	4.614830018	&		\\
ARHGEF7	&		&	-4.483844169	&		\\
PARVB	&		&	-4.455953222	&		\\
NSUN2	&		&	-4.366593041	&		\\
G6PC2	&		&	-4.34652349	&	*	\\
GCK	&		&	4.264128274	&	*	\\
ATXN7L1	&		&	4.221013512	&		\\
EVL	&		&	4.215670293	&		\\
FLJ46082	&		&	4.139703081	&		\\
RBMXL2	&		&	4.118967855	&		\\
STK40	&		&	4.080306367	&		\\
C4orf6	&		&	4.069766781	&		\\
PDGFRL	&		&	4.003454714	&		\\
RARB	&		&	-3.993516861	&		\\
ASAH1	&		&	-3.965690824	&		\\
TOM1L1	&		&	3.878585464	&		\\
KCNJ1	&		&	3.86017897	&		\\
PARD3	&		&	3.840674359	&		\\
SLC8A3	&		&	-3.704740594	&		\\
ZNF793	&		&	3.702417636	&		\\
SNX7	&		&	3.566606658	&		\\
NFATC2	&		&	3.532683466	&		\\
CNTN4	&		&	3.515837319	&		\\
MEST	&		&	3.437653132	&		\\
MAGI2	&		&	-2.782542891	&		\\
SPC25	&		&	2.587174283	&		\\
PARD3	&	ZAK	&	-2.465609064	&		\\
MTNR1B	&	PARD3	&	-1.803288774	&		\\
PARD3	&	MAGI2	&	0.547763814	&		\\
ZAK	&	MAGI2	&	-0.532970004	&		\\
ZAK	&	NFATC2	&	-0.462383518	&		\\
G6PC2	&	GCK	&	0.372561933	&		\\
ZAK	&		&	0.001314428	&		\\
\hline
\end{tabular}
\label{tab:4-5}
\end{table} 

\section{Summary}
In this chapter, we applied our proposed method to three real data sets in Framingham Heart Studies. For the first application of log IGE concentration, we used three different biological databases to construct W matrix KEGG pathway, GO biological process collection and HPRD protein-protein interactions database. Using the KEGG pathway database, we found 6 potentially biologically meaningful gene-by-gene interactions. The analyses using GO and HPRD databases added a couple additional such interactions. In all of the three analyses, we replicated 6 genes as main effects that were previously identified in \cite{Granada2012}. We also used stage-wise method implemented in Mendel to investigate the gene-by-gene interactions, but it didn\rq{}t find any potentially interesting interactions and only replicated two previously found genes in main effects.

Using the IGE dataset as an example, we also investigated the performance of our method under modest LD. We selected 1, 3, 5 SNPs per gene for SNPs on Chromosome 6 and compared the three analyses. The analyses with 3 SNPs and 5 SNPs per gene selected the same list of main effects and interactions on gene levels. We also compared the one-SNP/gene analysis and the 3\&5-SNP/gene analyses. In terms of the unique main effects and interactions, the effect of modest LD among SNPs did not seem to substantially affect the selection of terms (with Jaccard coefficient of 0.522). 

We applied the method to two more data sets in Framingham Heart Study, CRP serum levels and Fasting Glucose. We used only KEGG pathway database to construct W matrix, since we can see that in the first example using KEGG pathway is more effective in finding interesting interactions. The results of these two analyses also found some interactions involving pairs of previously identified genes.

\cleardoublepage

\chapter{Conclusion}
\label{chapter:Conclusion}
\thispagestyle{myheadings}


We developed a novel methodology to detect gene-by-gene interactions. We assessed the performance of this method under various scenarios in simulation, and compared it to a stage-wise competitor. The simulations showed that our method outperforms the competitor in finding true interactions, while maintaining about the same ability to detect main effects. Our method is robust to the inclusion of \lq{}noise interactions\rq{} between non-active SNPs. As showed in simulation, scaling up the number of predictors didn\rq{}t adversely affect our ability to detect interactions if the increased predictors are not involved in interactions. But it performs less well when \lq{}noise interactions\rq{} involve true active SNPs. Incorporating outside biological information as a network induced in the penalty term is an advantage of our method, in both reducing computing time and guiding selection of interactions. Additional simulations also showed that our method outperforms the simple association tests in detecting both interactions and main effects.

To extend our method to multi-cohort setting, we evaluated four procedures to conduct meta-analysis in simulations and found the approach that performs most closely to the mega-analysis which consists of merging individual level data are merged together and analyzing as one dataset. This procedure (B) is also more practical because it splits cohorts instead of splitting data within each cohort and thus simplifies the communication process among different study centers and reduce the possibility of making errors. 

Another advantage of Procedure B is that, it is relatively easy to implement multiple splits compared to other procedures. As stated in Chapter 3, all studies would be requested to perform variable selection analysis only once. The multiple splits of cohorts is conducted on the meta-analyst\rq{}s side and only affects the union lists of selected terms that may be generated by different assignments of studies in the first group. On the side of individual cohorts, they are asked to perform at most $K$ final model fittings based on their assignment of groups after the variable selection analysis, where $K$ is the number of splits. But for other procedures, for example, Procedure C, it is a much more complicated process. Each cohort would need to split their data $K$ times, perform variable selection $K$ times and final model fitting $K$ times. This is an error-prone process because the meta-analysis center has much less control over the analysis. We recommend using Procedure B when conducting meta-analysis using our method. 

Our meta-analysis approach is generalizable to other penalized regression methods, as the feature of our method that is used to develop the meta-analysis approach is no different than any other Lasso-based method.

We applied our proposed method to real datasets in Framingham Heart Study. As a typical example using KEGG pathway database to construct W matrix in the penalty, the IGE analysis showed that we found some potentially biologically interesting interactions and were able to identify many important main effect findings previously reported in publications as well. We explored the detection of interactions using two other outside biological sources GO and HPRD PPI database. These two analyses identified two additional interactions that are potentially interesting, one in pairs of previously identified genes and the other in the HLA region. We tested our method under modest LD using the IGE data and found that including multiple SNPs per gene did not substantially affect the result. We further applied our method to another two real data sets, CRP serum levels and fasting glucose. Using KEGG pathway database to construct W matrices for these two analyses, we found some interactions that may be interesting for these two phenotypes (one interaction in each analysis having genes both previously found).

\cleardoublepage


\begin{appendices}

\chapter{}
\thispagestyle{myheadings}

\section{Derivation of Model Fitting Algorithm} 

Our goal is to optimize the objective function
\begin{equation}
\begin{split}
f(\mathbf{\beta})=& \frac{1}{2} \left(\mathbf{Y}-\displaystyle\sum_{j=1}^{p}{\mathbf{X}_j\beta_j}-
\displaystyle\sum_{k>j}^{p}{\mathbf{X}_{jk}\beta_{jk}}\right)^T\left(\mathbf{Y}-\displaystyle\sum_{j=1}^{p}{\mathbf{X}_j\beta_j}-
\displaystyle\sum_{k>j}^{p}{\mathbf{X}_{jk}\beta_{jk}}\right)\\
& +\lambda_1\displaystyle\sum_{j=1}^{p}{(w_{jj}^2\|\mathbf{X}_j\beta_j\|^2+
\displaystyle\sum_{k: k \neq j}^{p}{w_{jk}^2
\|\mathbf{X}_{jk}\beta_{jk}\|^2})^{1/2}}+\lambda_2\displaystyle\sum_{j=1}^{p}
\displaystyle\sum_{k > j}^{p}{w_{jk}\|\mathbf{X}_{jk}\beta_{jk}\|} \label{eq:A1}
\end{split}
\end{equation}
in $\beta = \{\{\beta_j\},\{\beta_{jk}\}\}$,
where $\mathbf{Y}=(Y_1,\ldots,Y_n)^T$, $\mathbf{X}_j=(X_{1j},\ldots,X_{nj})^T$, and $\mathbf{X}_{jk} = (X_{1j}X_{1k}, \ldots, X_{nj}X_{nk})^T$.  Without loss of generality, we assume that $\mathbf{Y}$ and $\{\mathbf{X}_j\}_{j=1}^p$ have been centered about their mean and standardized to have unit norm (and, indeed, our computations have been done under this convention). However, all key formulas below are derived in fully un-standardized form, for consistency across variables, since standardization of $\mathbf{X}_j$ does not imply standardization of $\mathbf{X}_{jk}$ (i.e., $||\mathbf{X}_{jk}||\ne 1$).

To accomplish our optimization, we use a coordinate descent algorithm, which updates one element of $\beta$ at a time while holding all other elements fixed and cycles through all elements until convergence.  We describe the resulting one-dimensional optimizations separately for the main effects and for the interaction effects.

Consider the main effect coefficient $\beta_j$.  It is convenient to write the optimization  with
respect to this coefficient as
\begin{equation}
\min_{\beta_j} \frac{1}{2} \left(\mathbf{\tilde Y}_j - \mathbf{X}_j \beta_j\right)^T
                           \left(\mathbf{\tilde Y}_j - \mathbf{X}_j \beta_j\right)
        + \lambda_1\left(
        w_{jj}^2 ||\mathbf{X}_j\beta_j||^2 + c_j \right)^{1/2}
        + \mathcal{C}_j\enskip  \label{eq:A2} ,
\end{equation}
with $c_j=\sum_{k\ne j} w_{jk}^2 ||\mathbf{X}_{jk}\beta_{jk}||^2 $ and
\begin{equation}
\mathbf{\tilde Y}_j = \mathbf{Y} - \sum_{\ell\ne j}\mathbf{X}_{\ell}\tilde \beta_{\ell}
		- \sum_{\ell=1}^p \sum_{k>\ell} \mathbf{X}_{\ell k}
			\tilde \beta_{\ell k}\enskip , \label{eq:A3}
\end{equation}
where $\tilde \beta_{\ell}$ is the current value of $\beta_{\ell}$, $\tilde \beta_{\ell k}$ is the current value of $\beta_{\ell k}$, and $\mathcal{C}_j$ is all of the rest of the penalty that does not involve $\beta_j$.
Let $\hat{\beta}_j$ denote the OLS estimator from fitting a regression through the origin for $\tilde{\mathbf{Y}}_j$ on $\mathbf{X}_j$.  Having centered and rescaled our variables, it follows that  $\hat{\beta}_j=\mathbf{X}_j^T \tilde{\mathbf{Y}}_j$ and that (\ref{eq:A2}) may be re-expressed as
\begin{equation}
\min_{\beta_j} \frac{1}{2} ||\mathbf{X}_j||^2 \left(\hat\beta_j - \beta_j\right)^2
	+ \lambda_1\left( w_{jj}^2 ||\mathbf{X}_j\beta_j||^2 + c_j \right)^{1/2}
        + \mathcal{C}_j'\enskip . \label{eq:A4}  
\end{equation}
Differentiating the argument in (\ref{eq:A4}) with respect to $\beta_j$ and setting the result  to zero yields
\begin{equation}
\frac{\partial}{\partial\beta_j} =||\mathbf{X}_j||^2 \left( \beta_j - \hat\beta_j \right)
+ \frac{\lambda_1 w_{jj}^2 ||\mathbf{X}_j||^2 \beta_j}{(w_{jj}^2 ||\mathbf{X}_j||^2 \beta_j^2 + c_j)^{1/2}}
= 0\enskip . \label{eq:A5} 
\end{equation}

So our estimate $\tilde \beta_j$  of the main effect of interest, $\beta_j$, is the solution to
\begin{equation}
\beta_j\left(1 + \frac{\lambda_1 w_{jj}^2}{(w_{jj}^2 \mathbf{X}_j^T \mathbf{X}_j \beta_j^2 + c_j)^{1/2}}
\right) = \hat\beta_j\enskip . \label{eq:A6} 
\end{equation}
Hence our solution has the form $\tilde{\beta}_j=\alpha_j \hat{\beta}_j$, for some shrinkage parameter $\alpha_j \in [0,1]$, where $\alpha_j$  satisfies the equation
\begin{equation}
\alpha_j\left(1 + \frac{\lambda_1 w_{jj}^2}{(w_{jj}^2 \mathbf{X}_j^T \mathbf{X}_j \alpha_j^2 \hat\beta_j^2 + c_j)^{1/2}}
\right) = 1\enskip . \label{eq:A7} 
\end{equation}
When $c_j=0$, (\ref{eq:A6}) can be solved in closed form as 
\begin{equation*}
\tilde\beta_j = \hbox{sign}(\hat\beta_j)
\left(|\hat\beta_j| - \lambda_1 w_{jj}/(\mathbf{X}_j^T \mathbf{X}_j)^{1/2}\right)_{+} \enskip, 
\end{equation*}
  where $(\cdot)_{+}$ denotes `positive part'.

Similarly, to estimate interaction coefficients $\{\beta_{jk}\}$ we write the optimization with respect to, say, $\beta_{jk}$ as
\begin{equation}
\begin{split}
\min_{\beta_{jk}} & \frac{1}{2} \left(\mathbf{\tilde{Y}}_{jk}-\mathbf{X}_{jk}\beta_{jk}\right)^T\left(\mathbf{\tilde{Y}}_{jk}-\mathbf{X}_{jk}\beta_{jk}\right)\\
& +\lambda_1\displaystyle\sum_{\ell=1}^{p}{(w_{\ell \ell}^2\|\mathbf{X}_\ell \beta_\ell \|^2+
\displaystyle\sum_{m: m \neq \ell}^{p}{w_{\ell m}^2
\|\mathbf{X}_{\ell m}\beta_{\ell m}\|^2})^{1/2}}+\lambda_2\displaystyle\sum_{\ell=1}^{p}
\displaystyle\sum_{m > \ell}^{p}{w_{\ell m}\|\mathbf{X}_{\ell m}\beta_{\ell m}\|}
\end{split}   \label{eq:A8} 
\end{equation}
where
\begin{equation}
\mathbf{\tilde Y}_{jk} = \mathbf{Y} - \sum_{\ell=1}^{p}\mathbf{X}_{\ell}\tilde{\beta}_{\ell}
		- \sum_{m>\ell}\sum_{(\ell,m)\neq (j,k)} \mathbf{X}_{\ell m}\tilde {\beta}_{\ell m} \enskip . \label{eq:A9} 
\end{equation}
The optimization (\ref{eq:A8}) can be rewritten as
\begin{equation}
\begin{split}
\min_{\beta_{jk}} \frac{1}{2} \mathbf{X}_{jk}^T\mathbf{X}_{jk}(\hat{\beta}_{jk}-\beta_{jk})^2 & +  \lambda_1\displaystyle\sum_{\ell=1}^{p}{(w_{\ell \ell}^2\|\mathbf{X}_\ell \beta_\ell\|^2+
\displaystyle\sum_{m: m \neq \ell}^{p}{w_{\ell m}^2
\|\mathbf{X}_{\ell m}\beta_{\ell m}\|^2})^{1/2}} \\ & +\lambda_2\displaystyle\sum_{\ell=1}^{p}
\displaystyle\sum_{m > \ell}^{p}{w_{\ell m}\|\mathbf{X}_{\ell m}\beta_{\ell m}\|} 
\end{split}  \label{eq:A10} 
\end{equation}
where $\hat{\beta}_{jk}= {\mathbf{X}_{jk}^T \mathbf{\tilde{Y}}_{jk}} / {\mathbf{X}_{jk}^T\mathbf{X}_{jk}}$  is the OLS estimator from fitting a regression through the origin of $\tilde{\mathbf{Y}}_{jk}$ on $\mathbf{X}_{jk}$.

Reasoning as in the case of main effects, differentiating the argument in (\ref{eq:A10}) with respect to $\beta_{jk}$
and setting the result to zero yields
\begin{equation}
\begin{split}
\frac{\partial}{\partial\beta_{jk}}= & \mathbf{X}_{jk}^T\mathbf{X}_{jk}(\beta_{jk}-\hat{\beta}_{jk})\\
& +\lambda_1 \frac{w_{jk}^2\mathbf{X}_{jk}^T\mathbf{X}_{jk}\beta_{jk}}{(w_{jj}^2 \mathbf{X}_j^T \mathbf{X}_j \beta_j^2+\sum_{m \neq j}w_{jm}^2\mathbf{X}_{jm}^T\mathbf{X}_{jm}\beta_{jm}^2)^{1/2}} \\ 
& + \lambda_1 \frac{w_{jk}^2\mathbf{X}_{jk}^T\mathbf{X}_{jk}\beta_{jk}}{(w_{kk}^2 \mathbf{X}_k^T \mathbf{X}_k \beta_k^2+\sum_{m \neq k}w_{km}^2\mathbf{X}_{km}^T\mathbf{X}_{km}\beta_{km}^2)^{1/2}} \\ & +\lambda_2 w_{jk}(\mathbf{X}_{jk}^T\mathbf{X}_{jk})^{1/2}sign(\beta_{jk}) \\ & =0
\end{split} \enskip  \label{eq:A11} 
\end{equation}
which can be simplified to
\begin{equation}
\begin{split}
& \tilde{\beta}_{jk}\left(1+\lambda_1 w_{jk}^2[\frac{1}{(w_{jk}^2\mathbf{X}_{jk}^T\mathbf{X}_{jk}\tilde{\beta}_{jk}^2+c_1^{jk})^{1/2}}
+\frac{1}{(w_{kj}^2\mathbf{X}_{kj}^T\mathbf{X}_{kj}\tilde{\beta}_{kj}^2+c_2^{jk})^{1/2}}]\right) \\
& =sign(\hat{\beta}_{jk})(|\hat{\beta}_{jk}|-\lambda_2 w_{jk}(\mathbf{X}_{jk}^T\mathbf{X}_{jk})^{-1/2})_+
\end{split}  \label{eq:A12} 
\end{equation}
or
\begin{equation}
\begin{split}
& \alpha_{jk}\hat{\beta}_{jk}\left(1+\lambda_1 w_{jk}^2[\frac{1}{(w_{jk}^2\mathbf{X}_{jk}^T\mathbf{X}_{jk}\alpha_{jk}^2\hat{\beta}_{jk}^2+c_1^{jk})^{1/2}}
+\frac{1}{(w_{kj}^2\mathbf{X}_{kj}^T\mathbf{X}_{kj}\alpha_{jk}^2\hat{\beta}_{jk}^2+c_2^{jk})^{1/2}}]\right) \\
& =sign(\hat{\beta}_{jk})(|\hat{\beta}_{jk}|-\lambda_2 w_{jk}(\mathbf{X}_{jk}^T\mathbf{X}_{jk})^{-1/2})_+
\enskip  \end{split}  \label{eq:A13} 
\end{equation}
Here $c_1^{jk}=w_{jj}^2 \mathbf{X}_j^T \mathbf{X}_j \beta_j^2+\sum_{m \neq j,k} w_{jm}^2 \mathbf{X}_{jm}^T\mathbf{X}_{jm}\beta_{jm}^2$ and \\
$c_2^{jk}=w_{kk}^2 \mathbf{X}_k^T \mathbf{X}_k \beta_k^2+\sum_{m \neq k,j}w_{km}^2\mathbf{X}_{km}^T\mathbf{X}_{km}\beta_{km}^2$,
while $\alpha_{jk}\in [0,1]$ is a shrinkage parameter defining $\tilde{\beta}_{jk}=\alpha_{jk} \hat{\beta}_{jk}$.
Solving (\ref{eq:A13}) for $\alpha_{jk}$ yields $\tilde\beta_{jk}$.

Based on the calculations above, the coordinate descent algorithm optimizing (\ref{eq:A1}) is: 

\noindent {\bf{Algorithm:}}

\begin{enumerate}
\item {\bf initialize} $\tilde{\beta}_j$ and $\tilde{\beta}_{jk}$ for all $j,k \in \{1, \ldots, p\}$,
\item {\bf for} $j \in \{1, \ldots, p\}$, \\
        compute $\mathbf{\tilde Y}_j = \mathbf{Y} - \sum_{\ell\ne j}\mathbf{X}_{\ell}\tilde \beta_{\ell} - \sum_{\ell=1}^p \sum_{k>\ell} \mathbf{X}_{\ell k} \tilde \beta_{\ell k}$; \\
       compute $\hat{\beta}_j=\mathbf{X}_j^T \tilde{\mathbf{Y}}_j$; \\
       solve (\ref{eq:A7}) for shrinkage parameter $\alpha_j$;\\
      update $\tilde{\beta}_j=\alpha_j \hat\beta_j$.
\item{\bf for} $(j,k) \in 1 \leq j \leq k \leq p$, \\
        compute $\mathbf{\tilde Y}_{jk} = \mathbf{Y} - \sum_{\ell=1}^{p}\mathbf{X}_{\ell}\tilde{\beta}_{\ell} - \sum_{m>\ell}\sum_{(\ell,m)\neq (j,k)} \mathbf{X}_{\ell m}\tilde {\beta}_{\ell m}$; \\
        compute $\hat{\beta}_{jk}= {\mathbf{X}_{jk}^T \mathbf{\tilde{Y}}_{jk}} / {\mathbf{X}_{jk}^T\mathbf{X}_{jk}}$; \\
        solve (\ref{eq:A13}) for shrinkage parameter $\alpha_{jk}$;\\
        update $\tilde \beta_{jk} = \alpha_{jk} \hat\beta_{jk}$.
\item {\bf iterate} steps 2 and 3 until convergence.
\end{enumerate}

       
                  
\section{Optimizing the Algorithm} \label{app:app2}

Despite the fact that our algorithm replaces the original high-dimensional optimization in (\ref{eq:A1}) with an iteration over a collection of simpler, one-dimensional optimizations, and despite the expected efficiencies to be gained through use of an appropriately sparse matrix $W=[w_{jk}]$, application of the algorithm to datasets with large numbers of SNPs (e.g., millions or tens of millions) will still be computationally prohibitive.  Therefore, in order to accelerate the algorithm, we employ a `swindle', similar to that of Wu \textit{et al.} (2009).  The basic idea is to apply the algorithm to only a small, well-chosen subset of SNPs, thus estimating the coefficients $\beta_j$ and $\beta_{jk}$ for those SNPs, and to treat the coefficients of all other SNPs as zero.  This estimate is then checked as a solution to the full optimization problem, by direct evaluation of the Karush-Kuhn-Tucker (KKT) conditions corresponding to (\ref{eq:A1}).  If the KKT conditions are satisfied, we are done.  If not, the subset of SNPs upon which our algorithm is run is expanded, and we repeat the process.  The SNPs are ordered for inclusion in this process according to an appropriate scoring function.

In more detail, our approach is as follows.  We define an initial score for each predictor in the form
\begin{equation}
score_j=\left\vert \sum_{i=1}^n y_i x_{ij}\right\vert  \enskip , \label{eq:A14}
\end{equation}
 which captures the extent to which the phenotype $Y$ is correlated
with the $j$-th SNP (the data are assumed to be centered and scaled).  Recall that our smoothing
parameter $\lambda_1$ is chosen implicitly, by the user specifying a desired number $s$ of main effects to be included in the model.  For a given $s$, we extract those $k> s$ predictors with the largest scores, where $k$ is a multiple of $s$ (e.g., taken to be $k=10s$ in all of our implementations).  The algorithm described above is applied to this subset of $k$ predictors, yielding estimates of coefficients $\beta_j$ for $s$ main effects, as well as any interactions allowed to enter the model.  Coefficients of zero are assigned to any of the $k$ predictors that did not enter the model, to any interactions among the $k$ predictors that similarly did not enter the model, and to the main effects and interactions involving any predictors not among the $k$ on which the algorithm was applied.

Let
\begin{equation}
L(\beta) = \frac{1}{2}(\mathbf{Y}-\displaystyle\sum_{j=1}^{p}{\mathbf{X}_j\beta_j}-
\displaystyle\sum_{k>j}^{p}{\mathbf{X}_{jk}\beta_{jk}})^T(\mathbf{Y}-\displaystyle\sum_{j=1}^{p}{\mathbf{X}_j\beta_j}-
\displaystyle\sum_{k>j}^{p}{\mathbf{X}_{jk}\beta_{jk}}) \label{eq:A15}  
\end{equation}
 denote the least-squares term in (\ref{eq:A1}).  The coefficients obtained for the $k$ predictors used in the algorithm, both main effects and interactions, will satisfy the KKT conditions.   It remains to examine whether the zeros assigned as coefficients for all predictors not used in the algorithm satisfy the KKT conditions as well.  Direct calculation and examination of the resulting equations yields that the relevant KKT conditions for the optimization in (\ref{eq:A1}) are the following:
\begin{equation}
\beta_j = 0 \, \Rightarrow \, |\triangledown L(\beta)_j| \leq \lambda_1 w_{jj} (\mathbf{X}_j^T \mathbf{X}_j)^{1/2}, \qquad \textrm{given } \beta_{jk} = 0
\label{eq:A16} 
\end{equation}

\begin{equation}
\begin{split}
\beta_{jk} = 0 \, \Rightarrow \, & |\triangledown L(\beta)_{jk}| \leq (2\lambda_1+\lambda_2) w_{jk} (\mathbf{X}_{jk}^T\mathbf{X}_{jk})^{1/2}, \qquad \\
& \textrm{given } \beta_j=0, \, \beta_k=0, \, \beta_{jm}=0 \, \textrm{ for all } m \neq k \, \textrm{ and } \beta_{km} =0 \, \textrm{ for all } m \neq j
\end{split} \label{eq:A17} 
\end{equation}

\begin{equation}
\begin{split}
\beta_{jk} = 0 \, \Rightarrow \, & |\triangledown L(\beta)_{jk}| \leq (\lambda_1+\lambda_2) w_{jk} (\mathbf{X}_{jk}^T\mathbf{X}_{jk})^{1/2}, \qquad \\
& \textrm{given } \beta_j=0, \, \beta_{jm}=0 \, \textrm{ for all } m \neq k ; \,  \textrm{ or } \beta_k=0, \, \beta_{km} =0 \, \textrm{ for all } m \neq j
\end{split} \label{eq:A18} 
\end{equation}
Here $\triangledown L(\beta)_j$ and $\triangledown L(\beta)_{jk}$ are the derivatives of $L(\beta)$ respect to $\beta_j$ and $\beta_{jk}$, respectively.

We check conditions (\ref{eq:A16},\ref{eq:A17},\ref{eq:A18}) for the omitted predictors and interactions. If the conditions are satisfied we stop and the result is a global minimum. If any of the omitted predictors and interactions violate the conditions, we double $k$ and repeat the process until the conditions are satisfied. The KKT conditions are satisfied eventually since in the worst case $k$ equals the total number of predictors. However, in practice our experience is that with the choices of $s$ and $k$ used in our work (e.g., $s=20$ or $30$, and $k=10s$) the algorithm typically doubles few times if any.


\section{Selection of Tuning Parameters}

Our proposed method has two tuning parameters: $\lambda_1$ and $\lambda_2$.  Recall that $\lambda_1$ is set implicitly in our approach, through the user specifying a desired number of main effects to be kept in the model.   In addition, recall that we reparameterize $\lambda_2$ in terms of $\lambda_1$, as $\lambda_2 = c\lambda_1$.  Here we describe the derivation of a heuristic for choosing $c$ in a manner that is informed by the extent to which interactions can enter the model relative to main effects.

Consider a specific main effect, $\beta_j$, and an interaction effect, $\beta_{jk}$. In order to better focus the exposition squarely on $c$ and its connection to interactions, we will here explicitly utilize our assumption that $||\mathbf{X}_j||=1$, which allows us to drop all such terms from the expressions below.
By examining the equations for $\alpha_j$ (\ref{eq:A7}) and $\alpha_{jk}$ (\ref{eq:A13}), we can obtain thresholds of unshrunken estimates for the corresponding terms $X_j$ and $X_{jk}$ to enter the model. Specifically, for main effects,
\begin{equation}
\text{threshold for $X_j$ to enter} =
\begin{cases}
\lambda_1 w_{jj} & \text{, if $\beta_{jk} = 0$ for all $k$} \\
0     & \text{, otherwise}\enskip
\end{cases} \label{eq:A19} 
\end{equation}
whereas for interactions $\beta_{jk}$,
\begin{equation}
\begin{split}
& \text{threshold for $X_{jk}$ to enter} \\
& = \frac{w_{jk}}{(\mathbf{X}_{jk}^T\mathbf{X}_{jk})^{1/2}}
\begin{cases}
\lambda_2+ 2 \lambda_1 & \text{, if $\beta_j = \beta_k = 0$} \\
\lambda_2+ \lambda_1 & \text{, if $\beta_j \neq 0$ or $\beta_k \neq 0$ } \\
\lambda_2     & \text{, if $\beta_j \neq 0$ and $\beta_k \neq 0$}\enskip .
\end{cases}
\end{split} \label{eq:A20} 
\end{equation}
Having reparameterized $\lambda_2=c \lambda_1$, (\ref{eq:A20}) can be reexpressed as
\begin{equation}
\begin{split}
& \text{the threshold for $X_{jk}$ to enter} \\
& = \frac{\lambda_1 w_{jk}}{(\mathbf{X}_{jk}^T\mathbf{X}_{jk})^{1/2}}
\begin{cases}
c+ 2  & \text{, if $\beta_j = \beta_k = 0$} \\
c+ 1 & \text{, if $\beta_j \neq 0$ or $\beta_k \neq 0$ } \\
c     & \text{, if $\beta_j \neq 0$ and $\beta_k \neq 0$}\enskip .
\end{cases}
\end{split} \label{eq:A21} 
\end{equation}

Suppose that we parameterize by $r$  the relative difficulty of a main effect entering a model as compared to an interaction.  Then the expressions in  (\ref{eq:A19}) and (\ref{eq:A21}) suggest (ignoring the distinction between $c$, $c+1$, and $c+2$ in (\ref{eq:A21})) setting
\begin{equation}
r = \frac{\lambda_1 w_{jj}}{c\lambda_1 w_{jk} / (\mathbf{X}_{jk}^T\mathbf{X}_{jk})^{1/2}} = \frac{w_{jj}}{w_{jk}}\, \frac{(\mathbf{X}_{jk}^T\mathbf{X}_{jk})^{1/2}}{c} \enskip ,
\label{eq:A22} 
\end{equation}
and hence
\begin{equation}
c = \frac{w_{jj}}{w_{jk}}\, \frac{(\mathbf{X}_{jk}^T\mathbf{X}_{jk})^{1/2}}{r} \enskip .
\label{eq:value.of.c}
\end{equation}

The choice of $c$ therefore will vary inversely with the value $r$ specified by the user.  In addition, it will be influenced directly by the ratio of the values $w_{jj}$ and $w_{jk}$, i.e., the weights in the matrix $W$ associated with $\beta_j$ and $\beta_{jk}$ in the penalty function in the optimization (\ref{eq:A1}).  If  a $W$ matrix with ones down the diagonal is used, as we do, then $w_{jj}=1$, and so only the value $w_{jk}$ plays an explicit role in setting $c$.  If a choice is made to use a binary matrix $W$, as we do in our simulations, then $w_{jk}=1$.  In that case, the value of $c$ is driven purely by (a) the user-specified value of $r$, and (b) the value $(\mathbf{X}_{jk}^T\mathbf{X}_{jk})^{1/2}$.

In order to interpet this term, we reason as follows.  Recall that we have centered the $\mathbf{X}_j$ and rescaled them so that $\|\mathbf{X}_j\|=1$. Let $U_{j,i}$ be the original observation for individual $i$ on predictor $j$ after centering but before rescaling, then it has (approximately) mean zero and, say, variance $\sigma_j^2$.  Asymptotically, we have
\begin{equation}
X_{j,i} \approx \frac{1}{\sqrt{n} \sigma_j} U_{j,i} \enskip .
\label{eq:A24} 
\end{equation}
Since $\mathbf{X}_{j,k}$ is obtained through elementwise multiplication of $\mathbf{X}_j$ and $\mathbf{X}_k$,
\begin{eqnarray}
\mathbf{X}_{jk}^T\mathbf{X}_{jk} & \approx &
          \frac{1}{n\sigma_j \sigma_k} \mathbf{U}^T_{ji} \mathbf{U}_{jk} \nonumber \\
& &    \frac{1}{n\sigma_j\sigma_k} \sum_{i=1}^n U_{jk,i} U_{jk,i} \nonumber \\
& &    \frac{1}{n\sigma_j\sigma_k} \sum_{i=1}^n  U^2_{j,i} U^2_{k,i} \enskip .
\label{eq:A25} 
\end{eqnarray}
Now under an assumption of independence of predictors (i.e., independence of SNPs), we know that
$$E(U^2_{j,i} U^2_{k,i}) = E(U^2_{j,i}) E(U^2_{k,i}) = \sigma^2_j \sigma^2_k \enskip .$$
Assuming furthermore independent individuals, we have therefore by SLLN that $\mathbf{X}_{jk}^T\mathbf{X}_{jk} \longrightarrow \sigma_j\sigma_k$.
Therefore we have $(\mathbf{X}_{jk}^T\mathbf{X}_{jk})^{1/2} \approx (\sigma_j\sigma_k)^{1/2}$.

In other words, the term $(\mathbf{X}_{jk}^T\mathbf{X}_{jk})^{1/2}$ in (\ref{eq:value.of.c}) can be expected to behave (roughly) like the product of standard deviations of SNPs $j$ and $k$.  Because, for simplicity, we wish to use a single value $c$ for all interactions $\beta_{jk}$, in practice we use the mean or median (for our data, the difference was negligible) value of $(\mathbf{X}_{jk}^T\mathbf{X}_{jk})^{1/2}$ across SNP pairs for all pairs.  Writing the SNP variance as
$\sigma^2_j = 2\pi_j (1-\pi_j)$, where $\pi_j$ is the minor allele frequency for SNP $j$, the distribution of these terms can be
calculated efficiently by first calculating the $\sigma_j$ values and then their products (for all $j,k$ pairs with non-zero entry $w_{jk}$).

\end{appendices}
\cleardoublepage

\newpage
\singlespace
\bibliographystyle{apalike}


\bibliography{references}
\cleardoublepage


\end{document}